\documentclass[aps, preprint, longbibliography]{revtex4-1}
\usepackage{graphicx}
\usepackage{amsmath}

\begin{document}
	
	
	\title{An exploration of thermal counterflow in He II using particle tracking velocimetry}
	\author{Brian Mastracci}
	\author{Wei Guo}
	\email[Author to whom correspondence should be addressed. Electronic mail: ]{wguo@magnet.fsu.edu}
	\affiliation{National High Magnetic Field Laboratory, 1800 E Paul Dirac Dr., Tallahassee, FL 32310, USA}
	\affiliation{Department of Mechanical Engineering, Florida State University, 2525 Pottsdamer St., Tallahassee, FL 32310, USA}
	
	\date{\today}
	
	\begin{abstract}
		Flow visualization using PIV (particle image velocimetry) and particularly PTV (particle tracking velocimetry) has been applied to thermal counterflow in He II for nearly two decades now, but the results remain difficult to interpret because tracer particle motion can be influenced by both the normal fluid and superfluid components of He II as well as the quantized vortex tangle. For instance, in one early experiment it was observed (using PTV) that tracer particles move at the normal fluid velocity $v_n$, while in another it was observed (using PIV) that particles move at $v_n/2$. Besides the different visualization methods, the range of applied heat flux investigated by these experiments differed by an order of magnitude. To resolve this apparent discrepancy and explore the statistics of particle motion in thermal counterflow, we have applied the PTV method to a wide range of heat flux at a number of different fluid temperatures. In our analysis, we introduce a scheme for analyzing the velocity of particles presumably moving with the normal fluid separately from those presumably influenced by the quantized vortex tangle. Our results show that for lower heat flux there are two distinct peaks in the streamwise particle velocity PDF (probability density function), with one centered at the normal fluid velocity $v_n$ (named ``G2'' for convenience) while the other is centered near $v_n/2$ (``G1''). For higher heat flux there is a single peak centered near $v_n/2$ (``G3''). Using our separation scheme we show quantitatively that there is no size difference between the particles contributing to G1 and G2. We also show that non-classical features of the transverse particle velocity PDF arise entirely from G1, while the corresponding PDF for G2 exhibits the classical Gaussian form. The G2 transverse velocity fluctuation, backed up by second sound attenuation in decaying counterflow, suggests that large scale turbulence in the normal fluid is absent from the two peak region. We offer a brief discussion of the physical mechanisms that may be responsible for our observations, revealing that G1 velocity fluctuations may be linked to fluctuations of quantized vortex line velocity, and suggest a number of numerical simulations that may reveal the underlying physics in detail.
	\end{abstract}
	
	
	\maketitle
	
	\section{Introduction}
	\label{sec:Intro}
	Between the absolute zero temperature and the lambda transition temperature, $T_\lambda\approx2.17$~K, \textsuperscript{4}He exists in a superfluid phase called He II. Phenomenologically, a two fluid model, in which a superfluid condensate and normal fluid of thermal excitations coexist and are fully miscible~\cite{Tisza1938,Landau1941}, provides a useful description of the mechanics of He II. The superfluid component has temperature dependent density $\rho_s\left(T\right)$, and accounts for 100\% of the bulk fluid density $\rho$ at absolute zero ($\rho_s\left(T=0\right)=\rho$). It is inviscid, has no entropy, and its circulation is confined to quantized vortex lines, each with a single quantum of circulation $\kappa\approx10^{-7}$~m\textsuperscript{2}/s about a core approximately $\xi_0=0.1$~nm in diameter~\cite{Vinen2006}. On the other hand, the normal fluid component (density $\rho_n\left(T=T_{\lambda}\right)=\rho$) behaves in the same manner as a classical fluid, though interaction with quantized vortices gives rise to the non-classical force of mutual friction~\cite{VinenII,VinenIII}. Due to the strong influence of quantum effects, it has become common to refer to turbulence in He II as quantum turbulence~\cite{Vinen2006}.
	
	Perhaps the most common manifestation of quantum turbulence is thermal counterflow, the mechanism by which He II transports thermal energy. In the presence of a heat source, the normal fluid carries entropy away from the source, with velocity $\mathbf{v}_n$, while the superfluid moves toward it, with velocity $\mathbf{v}_s$, such that the overall mass flow is equal to zero, $\rho_n\mathbf{v}_n+\rho_s\mathbf{v}_s=0$~\cite{VanSciver2012}. It is important to recall that the two fluid components are fully miscible, such that these two velocity fields are interpenetrating; the counterflow currents are not spatially distinct as in a classical natural convection loop. In a simple one-dimensional case (e.g. counterflow through an insulated channel with a heater at one end), the normal fluid velocity is related to the magnitude of the heat flux, $q$, as 
	\begin{equation}
	v_n=\frac{q}{\rho{s}T}
	\end{equation}
	where $T$ denotes the fluid temperature and $s$ its specific entropy. The corresponding theoretical superfluid velocity is easily obtained by conservation of mass, $v_s=-v_n\rho_n/\rho_s$. As the heat flux increases, the counterflow velocity, $v_{ns}=v_n-v_s$, increases accordingly, and turbulence can develop in both fluid components~\cite{Guo2010,Marakov2015}. Superfluid turbulence manifests as a tangle of quantized vortex lines~\cite{VinenIII}, with the line length per unit volume $L$ approximated by 
	\begin{equation}
	\label{eq:lineDensity}
	L = \gamma^2\left(v_{ns}-v_0\right)^2
	\end{equation}
	where $\gamma$ is an experimentally determined temperature dependent parameter~\cite{Chase1962,Dimotakis1973,Childers1976,Martin1983,Babuin2012,Gao2017} and $v_0$ is a small critical counterflow velocity of approximately 2~mm/s~\cite{Babuin2012,Varga2015,Gao2017}.
	
	Flow visualization has become a popular tool for the study of thermal counterflow, with several different methods applied in the most recent two decades~\cite{Guo2014}. Visualization is accomplished by first seeding the fluid with small tracer particles, illuminating them with a light source (typically a laser beam shaped into a thin sheet), and capturing images of their location in the moving fluid~\cite{Raffel1998}. By analyzing particle displacement during the time interval between successive images, the flow velocity field can be deduced. 
	
	Analysis of counterflow visualization is particularly challenging because of the numerous factors that influence particle motion. Besides interactions with the normal fluid through viscous forces and the superfluid through inertial and added mass effects~\cite{Sergeev2009}, particles can become trapped on quantized vortices~\cite{Parks1966,Bewley2006}, which move at some velocity $\mathbf{v}_L\ne\mathbf{v}_s$. Furthermore, multiple numerical studies have shown that particles are not necessarily stationary on the vortices, but are thought to slide along the core due to a drag force exerted by the normal fluid~\cite{Kivotides2008c,Mineda2013}. A concrete understanding of particle motion in thermal counterflow has been the subject of numerous experimental, theoretical, and computational efforts.
	
	The first experiments by Zhang and Van Sciver made use of the particle image velocimetry (PIV) technique~\cite{Zhang2005}, in which a pair of images separated by a short time interval are segmented, and cross-correlation of the segments together with knowledge of the image separation time is sufficient to obtain the velocity vector for each segment~\cite{Raffel1998}. Zhang and Van Sciver studied counterflow in a vertical channel generated by a range of heat flux, $110\leq{q}\leq1370$~mW/cm\textsuperscript{2}, at a variety of temperatures, $1.62\leq{T}\leq2.00$~K. They found that for a one-dimensional counterflow, regardless of temperature or applied heat flux, the measured particle velocity, $v_p$, is approximately half of the theoretical normal fluid velocity: $v_p\approx{v_n}/2$.  According to the subsequent theory of Sergeev et al., the observed behavior can be explained by interactions between the particles and quantized vortex lines~\cite{Sergeev2006}.
	
	Other experimental investigations of particle motion in thermal counterflow have employed the particle tracking velocimetry (PTV) technique~\cite{Paoletti2008JPS,Chagovets2011,LaMantia2016,Kubo2017}, in which individual particle locations are tracked throughout a sequence of images. The results of Paoletti et al. show that some particle tracks correspond to the normal fluid motion, exhibiting relatively straight trajectories with mean particle velocity $v_p\approx{v_n}$ in the same direction as the heat current, while others show erratic behavior with net motion against the heat current~\cite{Paoletti2008JPS}. In this experiment the temperature range was $1.80\leq{T}\leq2.15$~K and the heat flux range was $13\leq{q}\leq90$~mW/cm\textsuperscript{2}, an order of magnitude less than that of Zhang and Van Sciver. The numerical work of Kivotides suggests two regimes of particle motion that are separated by the applied heat flux. The simulations show that when the vortex tangle is relatively dilute, as is the case when the applied heat flux is lower, particles have a relatively large mean free path through the tangle, with some traversing the entire observation volume at $v_n$ without interacting with vortices~\cite{Kivotides2008c}. On the other hand, when the tangle is relatively dense, particles cannot avoid interaction with vortices, and their mean velocity is lower than $v_n$~\cite{Kivotides2008b}. 
	
	Chagovets and Van Sciver used the PTV method intending to scan a parameter space covering that of the PIV experiment by Zhang and Van Sciver as well as the PTV experiment by Paoletti et al., thereby observing the transition between the two proposed flow regimes in a single experiment~\cite{Chagovets2011}. However, due to a hardware limitation, the heat flux range was limited to $7\leq{q}\leq100$~mW/cm\textsuperscript{2} at $1.55\leq{T}\leq2.00$~K~\cite{Chagovets2011}, which does not quite extend into the region probed by Zhang and Van Sciver. The results were nonetheless insightful, providing a discussion of the trapping of particles on quantized vortices and their subsequent dislocation, which presumably plays a role in the transition between the two regimes of particle motion~\cite{Chagovets2011}. Work has continued on classifying particle motion in thermal counterflow, with approaches focused on qualitative features of the particle trajectories~\cite{LaMantia2016} and analysis of particle motion as a function of their size~\cite{Kubo2017}.
	
	Another experimental approach to thermal counterflow that makes use of PTV is the analysis of transverse (i.e., perpendicular to the direction of normal fluid flow) particle velocity statistics. It has been shown that for both steady state~\cite{LaMantia2014a} and decaying~\cite{Paoletti2008} counterflow, the probability density function (PDF) for transverse particle velocity $u_p$ exhibits a Gaussian core with non-classical tails proportional to $\lvert{u_p}\rvert^{-3}$. In some cases, the tails are attributed to the motion of particles trapped on vortices that have just experienced a reconnection event~\cite{Paoletti2008}. Others point out that the tails can be predicted from the superfluid velocity field in the vicinity of a vortex core, without the need to consider vortex reconnection~\cite{LaMantia2014a,LaMantia2014b}. However, in light of numerical simulations that show particles suitably close to the vortex core have a tendency to become trapped rather than trace the superfluid velocity field~\cite{Barenghi2007,Kivotides2008}, this explanation is unlikely. Regardless, the tails have been shown to exist only when the probing time, $t_1$, is smaller than the average travel time between quantized vortex lines, $t_2=\ell/\left<v_p\right>$, where $\ell=L^{-1/2}$ represents the mean distance between vortex lines. When the ratio of these times, $\tau=t_1/t_2$, exceeds unity, the tails disappear and the PDF assumes the classical Gaussian form~\cite{LaMantia2014a}. This has been interpreted as an implication that counterflow turbulence behaves classically on large length scales~\cite{LaMantia2014a,LaMantia2014b}.
	
	It should be mentioned that flow visualization has been applied in He II for many purposes other than the study of particle motion. Some other investigations have been focused on counterflow~\cite{Zhang2005b,Chagovets2013,Duda2014} and forced flow~\cite{Chagovets2015} around cylinders; velocity profile in mechanically driven pipe flow~\cite{Xu2007}; dynamics of quantized vortices~\cite{Bewley2008,Paoletti2010,Fonda2014}; and flow induced by oscillating~\cite{Svancara2017} and towed grids~\cite{Mastracci2018}. More recently, a different approach to He II flow visualization has been introduced, making use of metastable He\textsubscript{2}* molecules as tracer particles~\cite{Gao2015}. Measurements of the turbulent normal fluid velocity with these particles have lead to non-classical forms of the second order transverse structure function~\cite{Marakov2015}, effective kinematic viscosity in decaying counterflow turbulence~\cite{Gao2016}, and the energy spectrum in a sustained thermal counterflow~\cite{Gao2017}. These measurements are free of the ambiguity associated with PIV and PTV methods since the He\textsubscript{2}* molecules strictly trace the normal fluid for temperatures above about 1~K~\cite{Gao2015}.
	
	In this paper we return to measurement of particle motion in thermal counterflow using PTV. However, we attempt to remove the particle motion ambiguity by analyzing particles that move with the normal fluid separately from those influenced by vortices. To our knowledge this approach has not yet been attempted. We were also successful at probing a wide range of applied heat flux, overlapping with the PIV experiments of Zhang and Van Sciver as well as with the PTV experiments of Paoletti et al. and Chagovets and Van Sciver. In Sect.~\ref{sec:Protocol}, we briefly describe our experimental protocol. The criteria for differentiating particle velocity measurements is covered in Sect.~\ref{sec:ResultsA}, and we showcase the results obtained for streamwise particle motion and transverse velocity statistics, respectively, in terms of the separated velocity measurements, in Sect.~\ref{sec:ResultsA} and Sect.~\ref{sec:ResultsB}. In Sect.~\ref{sec:Discussion} we offer a brief discussion of the physical mechanisms that may be responsible for our observations, and suggest a number of numerical simulations that may reveal the underlying physics in detail, before concluding in Sect.~\ref{sec:Conclusion}.
	
	\section{Experimental Protocol}
	\label{sec:Protocol}
\begin{figure}[b]
		\centering
		\includegraphics[]{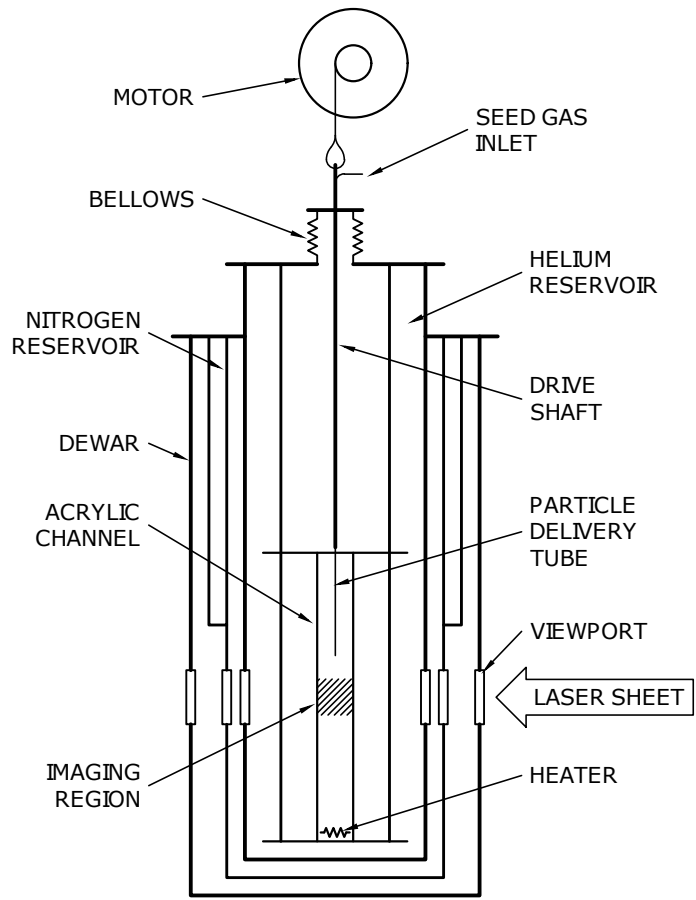}
		\caption{Simple illustration of the experimental apparatus (not to scale).\label{fig:setup}}
	\end{figure}

	Thermal counterflow is generated and contained inside a vertical flow channel which itself is immersed in the helium reservoir of a typical research cryostat with optical access. The channel, which was designed for visualization of both counterflow and towed grid turbulence in He II (see \cite{Mastracci2018} for details), is constructed from cast acrylic with a square cross section of 1.6~cm side length and measures 33~cm long. The bottom end is sealed with an array of evenly spaced surface mount resistors that occupy about 80\% of the channel cross section, such that the applied heat flux is distributed nearly uniformly. An illustration of the experimental apparatus is shown in Fig.~\ref{fig:setup}.
	
	For this work we make use of the PTV method. Tracer particles are formed by slowly introducing a gas mixture of 5\% D\textsubscript{2} gas (balance He) directly into He II~\cite{Fonda2016}. This seed gas is delivered via a tube that passes through the main linear drive shaft for the towed grid mechanism, and opens several centimeters above the imaging region. Typically, about 70\% of the resulting particles have diameters in the range 3--6~$\mu$m, as determined from their terminal velocity in quiescent He II~\cite{Mastracci2018}. Since particle size effectively sets the minimum spatial resolution~\cite{LaMantia2014b}, and we anticipate $\ell>10~\mu$m, these particles should be suitable for probing length scales both above and below the mean vortex line spacing. Before measurements begin, the particle delivery tube is retracted from the channel by raising the grid drive shaft (the mesh grid itself is removed for counterflow experiments). This prevents any flow structures that might develop upstream of the tube, a phenomenon known to occur in He II counterflow~\cite{Zhang2005b,Soulaine2017}, from interfering with the velocity field in the region of interest.
	
	A continuous wave laser, shaped into a thin sheet of approximately 16~mm height, provides illumination of the imaging plane in the geometric center of the channel. A high speed CMOS camera, triggered at various rates between 60 and 180 frames per second depending on the anticipated normal fluid velocity, captures sequences of several hundred images of the particles moving under the influence of counterflowing He II. Tracks are extracted from the sequence of images using an algorithm that is based on the feature point tracking routine of Sbalzarini and Koumoutsakos~\cite{Sbalzarini2005}, but that we have tailored for use with solidified tracer particles in He II. From the tracks, which are essentially lists of spatial coordinates separated by a known time interval, it is trivial to derive the particle velocity.
	
	Using this apparatus we have measured particle motion in steady state thermal counterflow at three temperatures, and a wide range of heat flux was applied at each temperature: $38\leq{q}\leq215$~mW/cm\textsuperscript{2} at $T=1.70$~K, $38\leq{q}\leq366$~mW/cm\textsuperscript{2} at $T=1.85$~K, and $17\leq{q}\leq481$~mW/cm\textsuperscript{2} at $T=2.00$~K. This parameter space substantially overlaps those of the existing PTV experiments ($13\leq{q}\leq90$~mW/cm\textsuperscript{2} at $1.80\leq{T}\leq2.15$~K~\cite{Paoletti2008JPS} and $7\leq{q}\leq100$~mW/cm\textsuperscript{2} at $1.55\leq{T}\leq2.00$~K~\cite{Chagovets2011}) and the original PIV experiment ($110\leq{q}\leq1370$~mW/cm\textsuperscript{2} at $1.62\leq{T}\leq2.00$~K~\cite{Zhang2005}). 
	
	\section{Streamwise particle behavior}
	\label{sec:ResultsA}
	
	\begin{figure}
		\centering
		\includegraphics[width=1\columnwidth]{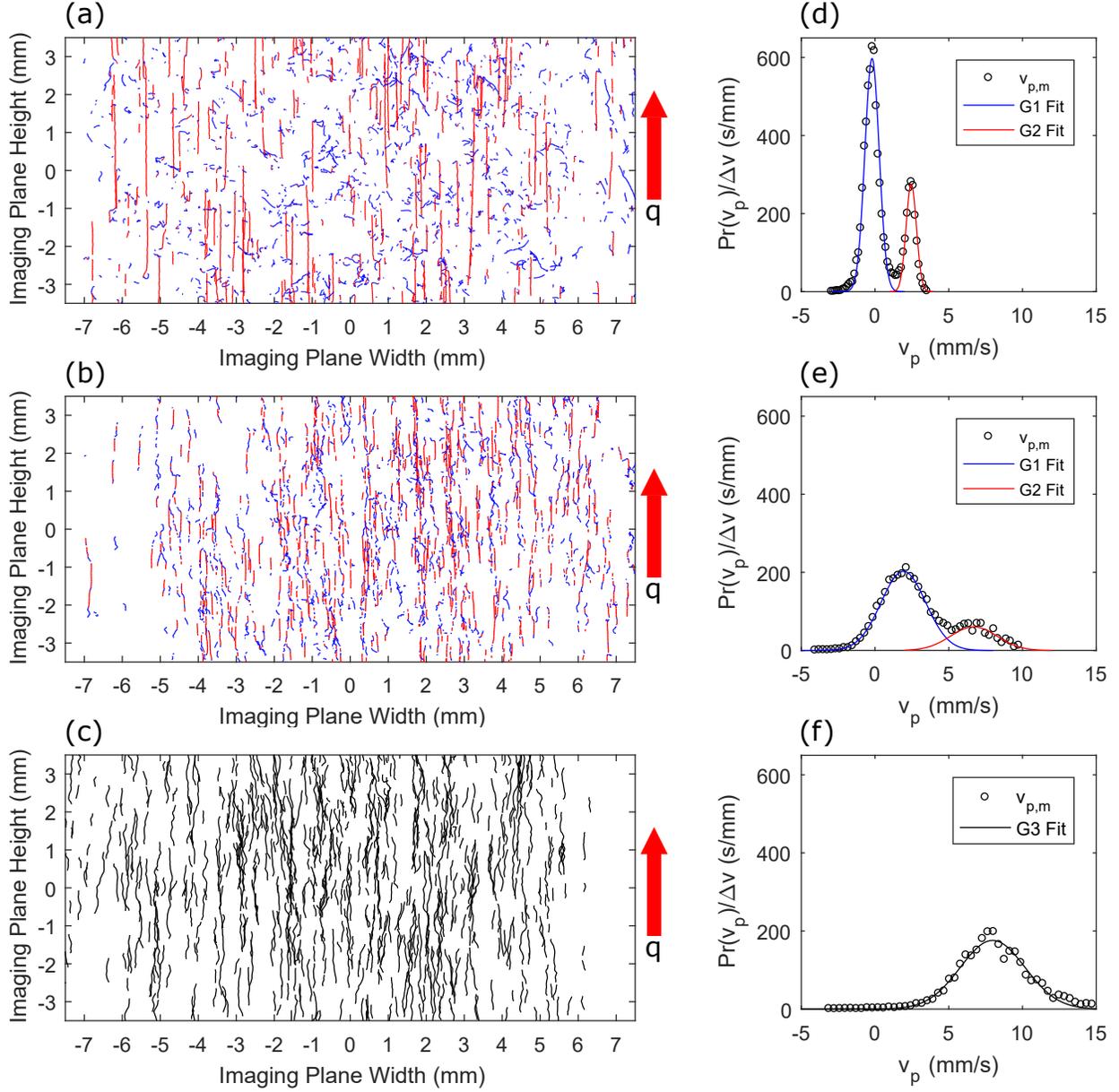}
		\caption{(a)-(c): Particle tracks extracted from videos for which $T=1.85$~K and $q=38$, $122,$ and $320$~mW/cm\textsuperscript{2}, respectively. (d)-(f): Corresponding particle velocity PDFs (streamwise direction). All six panels are color coded: blue indicates G1, red indicates G2, and black indicates G3. The direction of the imposed heat current is shown by the red arrows labeled ``$q$''.\label{fig:trackEvol}}
	\end{figure}
	
	Fig.~\ref{fig:trackEvol} shows some of the particle tracks observed at 1.85~K as well as the corresponding streamwise velocity PDFs. Though the particle tracks have the same structure as those shown in several previous studies~\cite{Paoletti2008JPS,Chagovets2011,LaMantia2016}, and it is well known that streamwise velocity PDFs, at least in the lower heat flux regime, exhibit two peaks~\cite{Paoletti2008,Paoletti2008JPS,LaMantia2012}, this figure showcases the novelty of our approach to the data analysis. We use the two-peak structure of the PDFs exemplified in Fig.~\ref{fig:trackEvol}(d) and (e) as the basis for analyzing the motion of particles moving with the normal fluid separately from those under the influence of the vortex tangle. Those moving with the normal fluid, whose velocity measurements contribute to the peak with higher mean value, we give the name ``Group 2'' or G2 for short. Those moving with the vortex tangle, whose velocity measurements contribute to the peak with lower mean value, we give the name ``Group 1'' or G1. For qualitative differentiation, we introduce the following criteria for deciding whether an instantaneous velocity sample represents G1 or G2 behavior. If the instantaneous velocity of a particle satisfies $v_p<\mu_2-2\sigma_2$, where $\mu_2$ and $\sigma_2$ are the mean and standard deviation, respectively, for a Gaussian curve fit to the G2 peak, we assume that it exhibits G1 behavior. Likewise, if $v_p>\mu_1+2\sigma_1$, we assume that it exhibits G2 behavior. In cases where $\mu_2-\mu_1>2\sigma_1+2\sigma_2$, i.e. the peaks are well separated, the criteria are reversed ($v_p<\mu_1+2\sigma_1$ counts as G1 and $v_p>\mu_2-2\sigma_2$ counts as G2) to prevent measurements falling in between the two peaks from counting toward both groups. As a result, the separation scheme generates ensembles of velocity measurements that represent G1 and G2. For brevity in the ensuing discussions, we use these names to refer interchangeably to the entire physical group of particles as well as the representative measurement ensemble. We define an additional group, G3, for the high heat flux regime. Since the streamwise velocity PDF exhibits just one peak for higher heat flux, as exemplified in Fig.~\ref{fig:trackEvol}(f), all of the measured velocity samples are representative of G3 behavior.
	
	The tracks and Gaussian fits of Fig.~\ref{fig:trackEvol} are color coded: G1 is shown in blue, G2 in red, and G3 in black. In the top row (Fig.~\ref{fig:trackEvol}(a) and (d)), which represents relatively low heat flux ($q=38$~mW/cm\textsuperscript{2}), the G2 tracks are long, straight, and oriented in the same direction as the heat current, while the G1 tracks meander and are randomly oriented. The corresponding peaks in the PDF are well defined (i.e., clearly separated from one another). In the middle row (Fig.~\ref{fig:trackEvol}(b) and (e)), which represents moderate heat flux ($q=122$~mW/cm\textsuperscript{2}), the G2 tracks are still straight and vertically oriented, but are frequently interrupted by short G1 segments. This likely represents the trapping of particles by quantized vortices, and the subsequent dislocation of the particle that stems from the increased normal fluid drag force~\cite{Kivotides2008c,Chagovets2011}. As a whole, the G1 tracks move in the same direction as the heat current, though in a slower and considerably less undeviating fashion than the G2 tracks. A positive shift in the mean value of both peaks can be observed in Fig.~\ref{fig:trackEvol}(e), and the peaks are less well defined, appearing to merge together. Though the evolution of particle velocity as a function of applied heat flux in this regime has been thoroughly discussed by Chagovets and Van Sciver, their assumption that the two group behavior continues indefinitely does not appear to be correct~\cite{Chagovets2011}. The bottom row (Fig.~\ref{fig:trackEvol}(c) and (f)), representing higher heat current ($q = 320$~mW/cm\textsuperscript{2}), shows that G3 tracks are all oriented in the same direction as the heat current but exhibit significant transverse motion, and their PDF exhibits only one peak. 

\begin{figure}[b]
		\centering
		\includegraphics[]{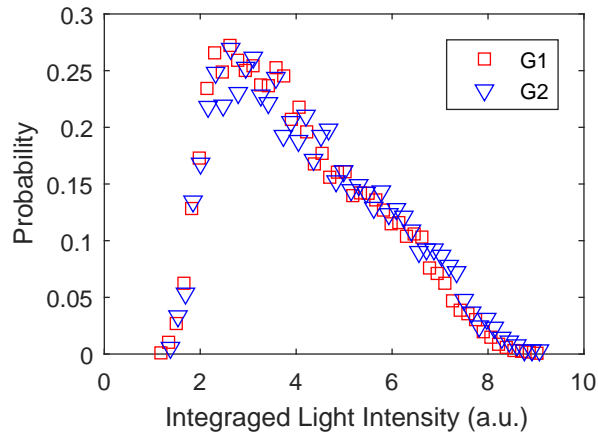}
		\caption{Probability distribution for the size (indicated by integrated light intensity) of particles contributing to G1 and G2. The example shown applies to the case where $T=1.85$~K and $q=38$~mW/cm\textsuperscript{2}.\label{fig:size}}
	\end{figure}

	Naturally, a question arises about what causes particles to move under the influence of the normal fluid or the vortex tangle. Many discussions on the behavior of particles in thermal counterflow mention particle size~\cite{Paoletti2008JPS,Kivotides2008,Chagovets2011,Kubo2017}. Using our separation scheme, we computed the PDF for integrated light intensity, or the sum of pixel values in the neighborhood of the particle image, which is used as a substitute for particle size since the latter cannot be accurately measured for a moving particle. Fig.~\ref{fig:size} shows that the PDF for G1 and G2 are nearly identical across the full range of observed particle size. This suggests that for solidified tracer particles in the size range produced by our seeding system, trapping probability is not influenced by particle size, though the observed range is quite small. We do not presently have an explanation for this.
	
	An additional consideration is that at the beginning of the image acquisition, particles are either trapped or untrapped, and whether the G2 particles become trapped during the acquisition period depends primarily on their mean free path through the vortex tangle. As a very simple estimation, we assume that a particle will become trapped if the volume traversed by its trapping cross section contains a line segment comparable in length to the particle diameter. We use for the trapping cross section the two-dimensional projection of the particle, $\pi{d_p}^2/4$. The volume traversed by the cross section is then $s\pi{d_p}^2/4$, where $s$ denotes the mean free path. Multiplication by $L$ gives the vortex line length within this volume, and as per our estimation, the result must be less than $d_p$ for the particle to remain untrapped:
	\begin{equation}
	\frac{\pi}{4}{d_p}^2sL\le{d_p}
	\end{equation}
	A simple representation for the mean free path is then 
	\begin{equation}
	\label{eq:mfp}
	s\lesssim\frac{4}{\pi{d_p}L}
	\end{equation}
	As examples we consider the cases shown in Fig.~\ref{fig:trackEvol}(a) and (b) for which we estimate $L\approx743$~cm\textsuperscript{-2} and $L\approx26913$~cm\textsuperscript{-2}, respectively, using the value for the $\gamma$ parameter reported by Gao et al.~\cite{Gao2017}. For particles with diameter 4~$\mu$m, the estimated mean free path is about 4~cm for case (a). This exceeds the dimensions of the imaging region, and the G2 tracks are quite long and often terminate when the particle leaves the imaging plane instead of with a transition to G1 behavior, which would indicate trapping. For case (b) the mean free path is about 0.1~cm, and it can be seen the length of many G2 tracks is roughly 1~mm, and the tracks often terminate in a trapping event. Though this simple estimation is reasonably accurate, a proper determination of the mean free path requires complex numerical simulations, taking into account the complicated dynamics of He II, such as Kelvin waves on quantized vortices, drag force exerted by the normal fluid, and relative motion of the particles and vortex tangle. Similar simulations by Kivotides indeed show that when the vortex tangle is relatively sparse, particles can move a significant distance (in some cases throughout the entire computational domain) without interacting with vortices~\cite{Kivotides2008c}, but when the tangle is relatively dense the particles experience constant interaction with the tangle~\cite{Kivotides2008b}.
	
	\begin{figure}[]
		\centering
		\includegraphics[]{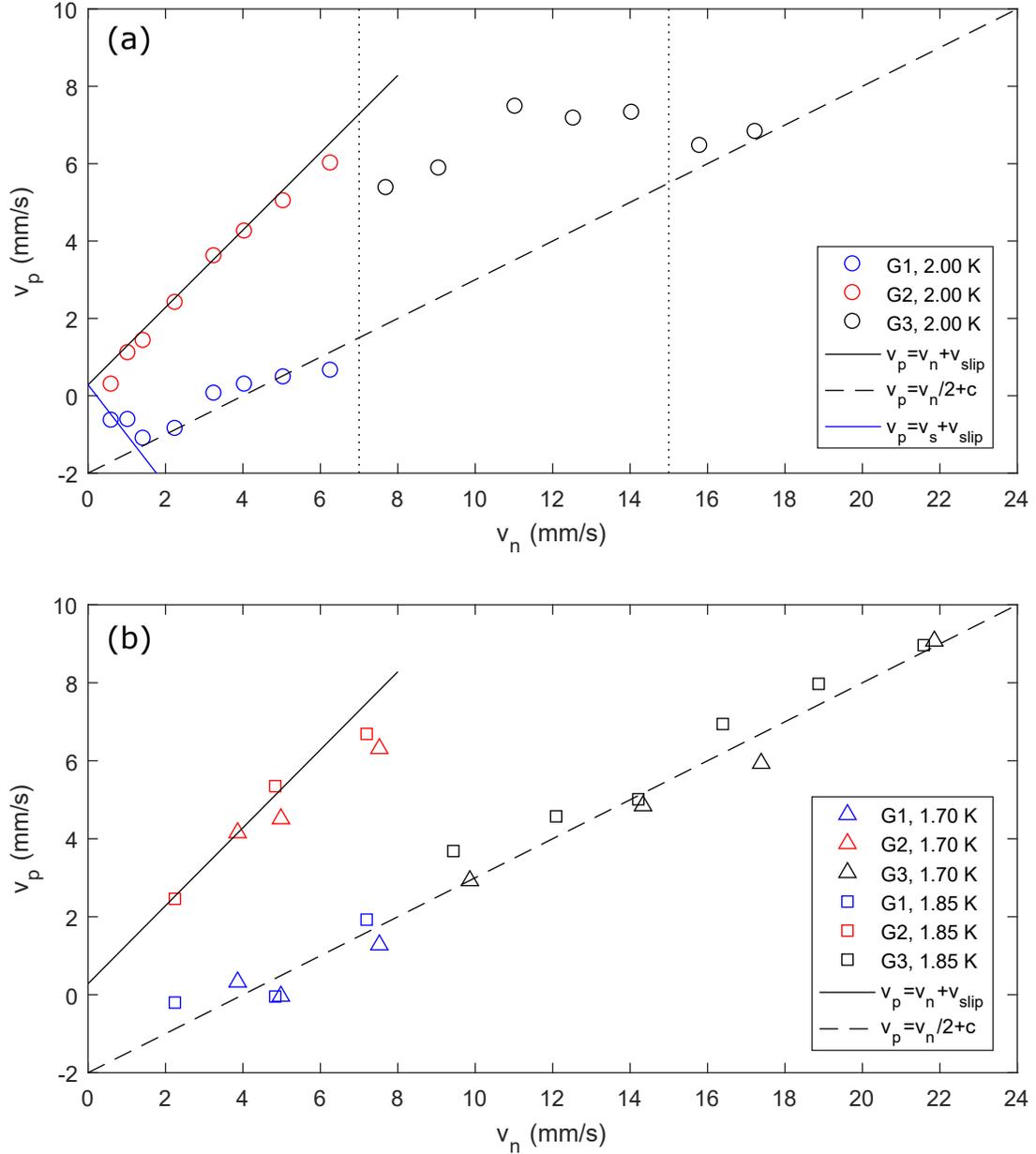}
		\caption{Measured particle velocity $v_p$ as a function of the theoretical normal fluid velocity $v_n$ for (a) $T=2.00$~K and (b) $T=1.85$~K and $T=1.70$~K.\label{fig:vpvsvn}}
	\end{figure}

	For consistency with the existing experimental literature~\cite{Zhang2005,Paoletti2008JPS,Chagovets2011,LaMantia2012} we show $v_p$ as a function of $v_n$ for each point in the parameter space in Fig.~\ref{fig:vpvsvn}. The top panel shows results only for $T=2.00$~K and the bottom panel shows results for the other two temperatures. $v_p$ is represented by the mean value of Gaussians fit to the streamwise velocity PDFs (as in the examples of Fig.~\ref{fig:trackEvol}(d)-(f)). In all cases, $v_p$ for the G2 peak is approximately equal to $v_n+v_{slip}$, where $v_{slip}$ is the velocity offset caused by non-neutral density of the particles. This trend is indicated by the solid black line. Mean velocity of the G1 peak, for very small heat flux, is similar to the superfluid velocity with the same correction factor, $v_s+v_{slip}$, indicated by the blue line. This behavior is expected for low counterflow velocities since the superfluid carries the vortex tangle, on average, at $v_s$~\cite{Wang1987,Donnelly1991}, and it has been demonstrated in recent visualization experiments~\cite{Paoletti2008JPS,Chagovets2011} and numerical simulations~\cite{Mineda2013}. As the heat flux increases and mutual friction begins to affect the vortex tangle, the G1 velocity departs from $v_s+v_{slip}$ and instead corresponds to $v_n/2+c$, indicated by the dashed black line, where $c$ is an offset of about 2~mm/s. At 2.00~K, as the heat flux continues to increase, the single peak PDF structure appears, with the mean value beginning from some value between $v_n$ and $v_n/2+c$ and eventually settling at the latter. This transition region occurs between normal fluid velocities of roughly 7 and 15~mm/s, as indicated by the vertical dotted lines in Fig.~\ref{fig:vpvsvn}(a). However, for 1.85 and 1.70~K, this transition appears to be absent, with the mean value of G3 PDFs immediately collapsing onto the $v_n/2+c$ trendline when $v_n$ exceeds 7 or 8~mm/s. We suspect that this results from a limitation of our imaging system; it seems that we are not able to resolve particles moving faster than about 9~mm/s. Since the dynamic viscosity of the normal fluid is smaller at these temperatures than at 2.00~K~\cite{Donnelly1998}, it makes sense that the critical drag force preventing particles from remaining trapped on vortices~\cite{Chagovets2011} is surpassed at higher values of $v_n$ that may be beyond those that we can resolve. In this case, the G3 data shown in Fig.~\ref{fig:vpvsvn} would in fact be miscategorized G1 data. As an additional note, small values of $v_{ns}$ were not observed at these temperatures, so the transition of G1 velocity from $v_s+v_{slip}$ to $v_n/2+c$ does not appear either.
	
	These observations are consistent with the existing literature on experimental measurements of particle motion in thermal counterflow: when the applied heat flux is lower, particles can be observed moving at approximately $v_n$~\cite{Paoletti2008JPS}, and when it is higher, particles can be observed moving at approximately $v_n/2$~\cite{Zhang2005}. However, we believe this is the first time that one experiment yields both observations, experimentally confirming a long-held theory that the early discrepancy was a matter of different flow regimes occurring for different ranges of the applied heat flux~\cite{Sergeev2009}.
	
	\section{Particle velocity statistics}
	\label{sec:ResultsB}
	Statistical analysis of particle motion in thermal counterflow using PTV is typically focused on the evolution of transverse particle velocity or acceleration PDFs with changing temperature, heat flux, or most commonly, probing time scale~\cite{LaMantia2013,LaMantia2014a,LaMantia2014b,LaMantia2016b}. In these analyses the statistical sample consists of all of the detected particles. This approach raises some concern when one considers the vastly different characteristics of the transverse motion exhibited by the G1 and G2 tracks in Fig.~\ref{fig:trackEvol}(a), and to a lesser extent in Fig.~\ref{fig:trackEvol}(b). Our analysis of the transverse particle velocity for G1 and G2 shows that some information is indeed missed when the two groups are not considered separately.
	
	First we note that, within each group, the streamwise and transverse velocity components are uncorrelated. In other words, the samples taken from any slice of the streamwise PDF will accurately represent the entire transverse distribution (provided the extracted sample size is large enough), and vice versa. This is important to the success of our separation scheme since the streamwise velocity PDFs for G1 and G2 end up quite lopsided. However, as will be seen in the figures of this section, the transverse velocity PDFs are sufficiently resolved.
	
	\begin{figure}[]
		\centering
		\includegraphics[width=0.9\columnwidth]{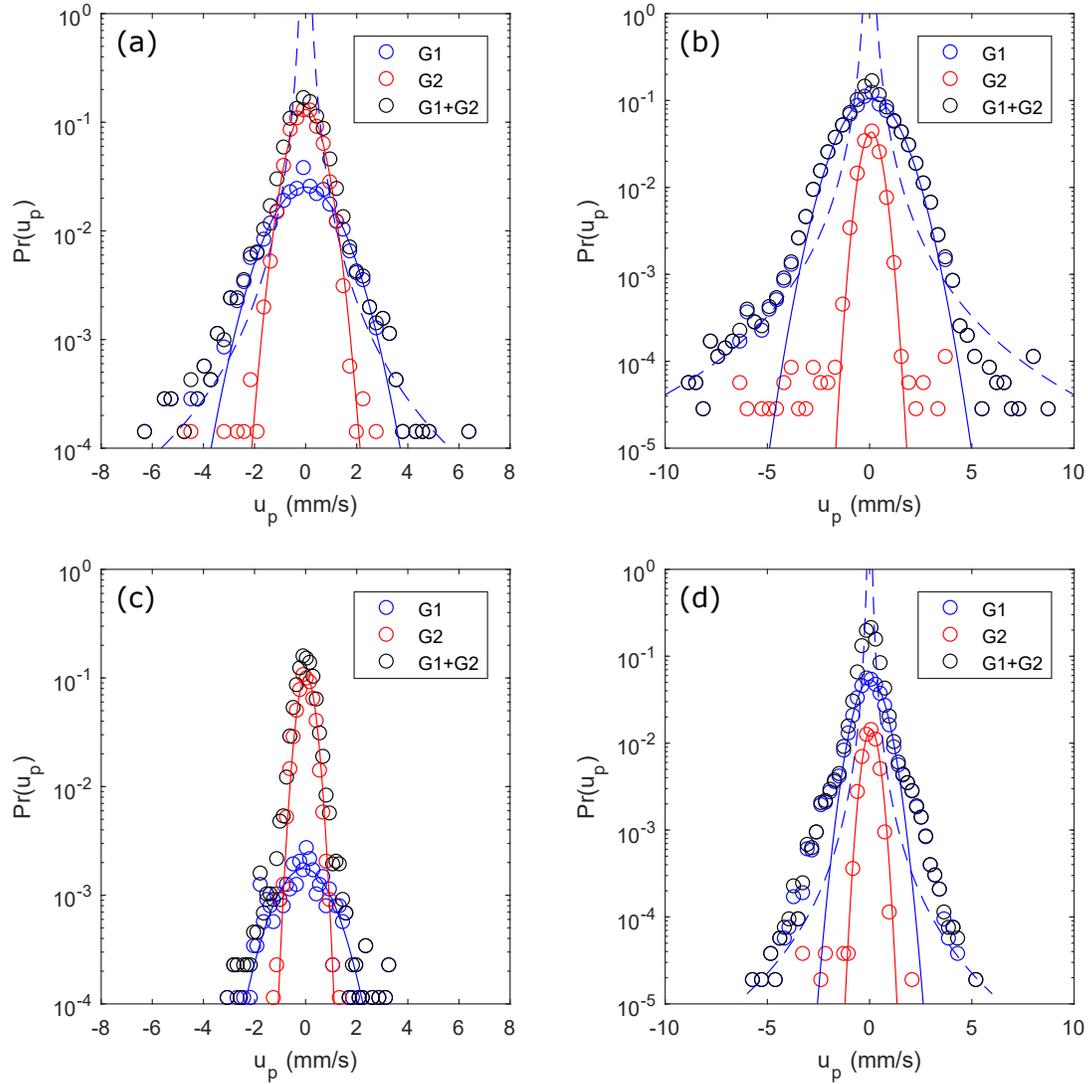}
		\caption{Probability distributions for the measured transverse particle velocity at $T = 2.00$~K for the cases where (a) $q = 91$~mW/cm\textsuperscript{2} and (b) $q = 113$~mW/cm\textsuperscript{2}. The solid lines represent Gaussian fits to the distributions and the dashed line represents a power law curve proportional to $\left|u_p\right|^{-3}$. Panels (c) and (d) show the same data but the minimum probing length has been changed to $2\ell$.\label{fig:pdfX}}
	\end{figure}
	
	Fig.~\ref{fig:pdfX} shows G1, G2, and combined (G1+G2) transverse velocity PDFs. In Fig.~\ref{fig:pdfX}(a) ($q=91$~mW/cm\textsuperscript{2}) the G2 sample size exceeds the G1 sample size. In Fig.~\ref{fig:pdfX}(b) ($q=113$~mW/cm\textsuperscript{2}) the opposite is true. To show the relative contributions of G1 and G2 to the combined PDF, the normalization is $Pr_{g,i}=n_{g,i}/N$, where $n_{g,i}$ is the number of samples in the $i^{th}$ bin for group $g$, and $N$ is the total number of combined G1 and G2 samples. It is clear that, regardless of the relative sample size, G1 dominates the tail region of the combined PDF. Though the G2 PDFs appear to have some structure at the ends, it is not coherent and occurs with probability at least an order of magnitude less than the corresponding G1 contribution. This is likely due to a small number of misclassified velocity measurements; those with streamwise component more than two standard deviations outside the group mean can be potentially placed in the wrong group. This effect could be confirmed by inspecting the location of these specific velocity samples in the particle tracks, and judging based on the local geometry whether they truly belong to G1 or G2.

	We also observe that the Gaussian core of the G1 PDF is substantially wider than the G2 PDF. This becomes of consequence when the G2 sample size is larger, as in Fig.~\ref{fig:pdfX}(a). As a result, the combined PDF may be broken into three regions. The tip region is Gaussian and due primarily to the G2 PDF. The middle region is due to the combined G2 PDF and Gaussian core of the G1 PDF, and has a different mathematical description than the tip region. The tail region exhibits the $\left|u_p\right|^{-3}$ power law behavior due exclusively to the G1 PDF tails. If the combined PDF is considered alone, it is possible to mistake the middle region for the beginning of the power law tails, leading to incorrect conclusions about the particle velocity statistics.
	
	Fig.~\ref{fig:pdfX}(c) and (d) show the same data as Fig.~\ref{fig:pdfX}(a) and (b), respectively, except the minimum probing length scale has been increased to $2\ell$, twice the mean vortex line spacing. It is important to note that this differs from the approach described in the existing literature~\cite{LaMantia2014a}, which is an adjustment of the probing time scale. The latter is achieved by using every other, or every third, etc., position measurement along a particle track to calculate the velocity, simulating a reduction in the image acquisition rate~\cite{LaMantia2014a}. Alternatively, a true adjustment of the minimum probing length can be accomplished by discarding position measurements only if they are not sufficiently separated from the previous location in the trajectory. Velocity samples are then computed as $\mathbf{v}_i=\left(\mathbf{x}_{i+j}-\mathbf{x}_i\right)/j\Delta{t}$, where $i$ represents the $i^{th}$ position along a track and $j$ represents the number of subsequent points to skip such that $\lVert\mathbf{x}_{i+j}-\mathbf{x}_i\rVert$ exceeds the desired minimum length scale. Increasing the probing length in this manner results in a drastic reduction of the number of G1 samples, since the mean G1 velocity is small compared to the mean G2 velocity. For a fixed sample size, the PDF tails are quenched since many of the measurements contributing to them are discarded. This is apparent in Fig.~\ref{fig:pdfX}(c), where the probability of observing a particle with G1 velocity is drastically reduced, and the extents of the PDF do not resemble the power law curve. The tails of Fig.~\ref{fig:pdfX}(d) are more or less eliminated as well, though the effect is not as obvious since the G1 sample size for this case was considerably larger. 
	
\begin{figure}[b]
		\centering
		\includegraphics[width=1\columnwidth]{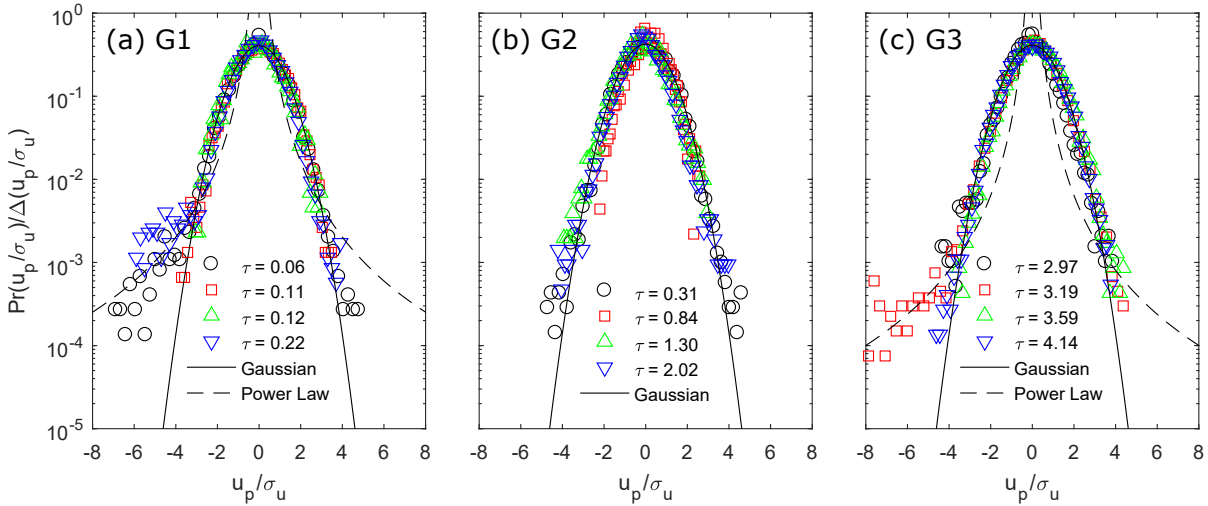}
		\caption{Normalized probability distribution for the measured transverse velocity of particles contributing to (a) G1, (b) G2, and (c) G3 at $T=2.00$~K. Several different values of the ratio $\tau$ are shown for each group. Gaussian curves ($A\exp\left(-\left(u_p/\sigma_u\right)^2/2\right)$) are fit to the entire PDF and power law curves ($A\left|u_p/\sigma_u\right|^{-3}$) are fit only to the tail regions, defined as more than $4\sigma_u$ from the center.\label{fig:pdfTau}}
	\end{figure}

	An alternative way to present the data is shown in Fig.~\ref{fig:pdfTau}, which contains several transverse velocity PDFs for each of G1, G2, and G3. Several curves with different values of the non-dimensional time $\tau$ are shown in each case. In the same manner as the existing literature, we define $\tau=t_1/t_2$, where $t_1$ is the time elapsed between successive images and $t_2=\ell/\left<v_p\right>$~\cite{LaMantia2014a}, except that in this case $\left<v_p\right>$ is computed for each group instead of for all of the detected particles. Defined in this way, $t_2$ represents the average time for a particle of each respective group to traverse the intervortex distance.
	
	A Gaussian form is evident in the core of all PDFs, regardless of group or probing time scale, as indicated by the solid black curve in all three panels ($A\exp\left(-\left(u_p/\sigma_u\right)^2\right)$, where $A$ is a constant). This is consistent with the existing literature~\cite{LaMantia2014a,LaMantia2014b}. For G1, shown in Fig.~\ref{fig:pdfTau}(a), a power law curve ($A\lvert{u_p}/\sigma_u\rvert^{-3}$, where $A$ is a constant), indicated by the dashed curve, can be drawn through the tail region. We define the tails as the data that falls more than $4\sigma_u$ from the center of the PDF. While the size of this data set is not sufficient to resolve extended tails, deviation from the Gaussian profile is clear, and the nondimensional time is less than unity for all of the cases shown, indicating that the probing time is smaller than the average intervortex travel time. According to the existing literature, these are the correct conditions for power law tails~\cite{LaMantia2014a,LaMantia2014b}.
	
	The G2 PDFs of Fig~\ref{fig:pdfTau}(b) present a different picture: all cases show purely Gaussian form, even though $\tau<1$ for some data and $\tau>1$ for others. This is likely because G2 consists of velocity measurements contributing to the part of the streamwise PDF that is normally attributed to particles moving with the normal fluid~\cite{Paoletti2008JPS,Chagovets2011}. Since the normal component behaves more or less classically, it makes sense for the normal fluid transverse velocity PDF to have the same Gaussian form that a classical fluid PDF would have.
	
	Despite the relatively long probing times ($\tau\gtrsim3$) for G3 PDFs, shown in Fig.~\ref{fig:pdfTau}(c), deviation from the Gaussian core can be observed in one case. As with G1, a power law curve can be drawn through this tail structure. Previous experimental results suggest that the PDF should have Gaussian form if $\tau>1$, but those investigations did not include the high heat flux G3 region~\cite{LaMantia2014a}. Indeed, little is known about the novel form of turbulence that exists in this high heat flux region, where both the normal fluid~\cite{Marakov2015} and superfluid can become turbulent.
    
    \begin{figure}[]
        \centering
        \includegraphics[width=\columnwidth]{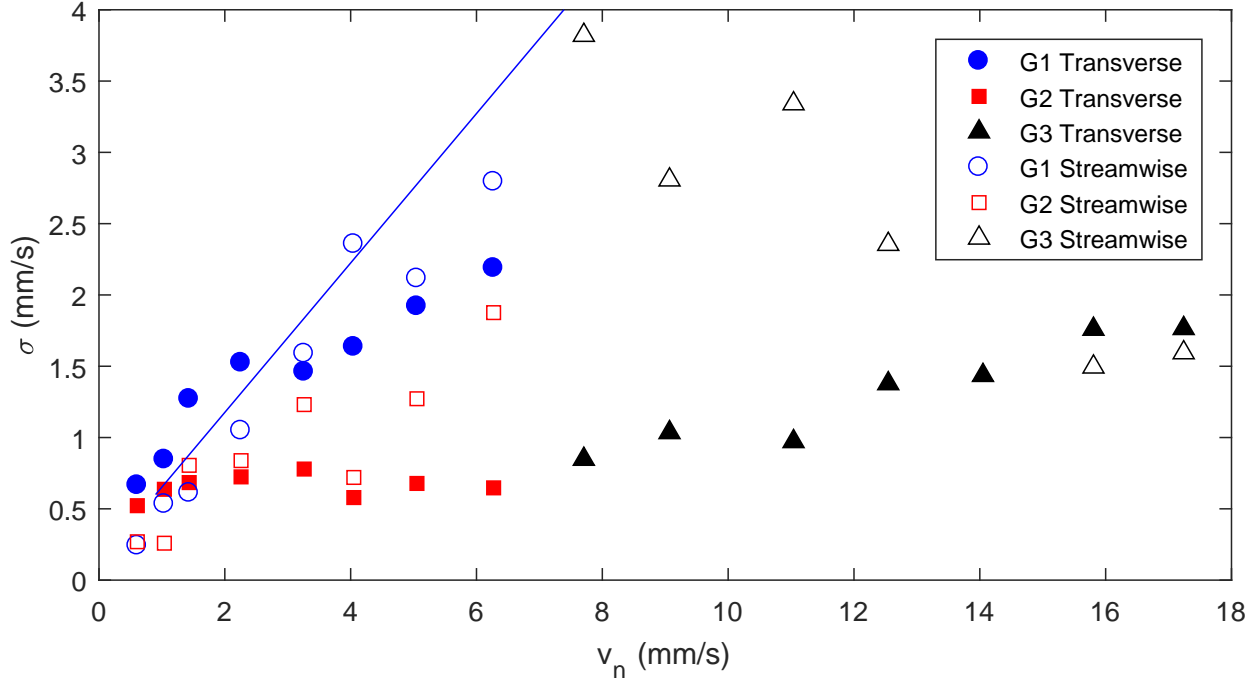}
        \caption{Standard deviation of the measured streamwise and transverse particle velocity for G1, G2, and G3 at $T = 2.00$~K. The solid blue line represents the vortex line velocity fluctuation and will be discussed in the next section. Note that the probing scale for these data is not constant.\label{fig:std}}
    \end{figure}
	
	We show in Fig.~\ref{fig:std} the standard deviation of the Gaussians fit to all streamwise and transverse velocity data obtained at 2.00~K. As in Sect.~\ref{sec:ResultsA}, data for G1 is shown in blue, G2 in red, and G3 in black. At first glance, the figure adds weight to the importance of the separation scheme, particularly for the analysis of transverse velocity statistics, since a clear divergence between G1 and G2 transverse velocity fluctuation is evident as the heat flux increases. Closer inspection reveals some additional, more subtle, observations. The measured transverse velocity standard deviation $\sigma_u$ for G2 is fairly constant throughout the range of applied heat flux at 2.00~K; the velocity fluctuation does not increase noticeably until the transition to G3. This suggests that the normal fluid may not be turbulent in the two peak region, which is further supported by a brief test of decaying thermal counterflow at 1.70~K in the two-peak region (G1 and G2, $q=50$~mW/cm\textsuperscript{2}) and the single peak region (G3, $q = 193$~mW/cm\textsuperscript{2}). In the single peak region the line density decays briefly as $L\propto{t}^{-1}$ before transitioning to $L\propto{t}^{-3/2}$, whereas in the two peak region it follows $L\propto{t}^{-1}$ throughout the entire decay. Gao et al. have shown that the former behavior corresponds to decay from a steady state counterflow in which large scale turbulence exists in the normal fluid, while the latter decay behavior occurs when normal fluid turbulence is absent~\cite{Gao2015JETP}. It is worthwhile to note that if we scale the normal fluid turbulence transition heat flux reported by Gao et al. to the wider channel used for our experiment, we obtain a heat flux slightly smaller than that at which the two peak structure disappears. However, the transition to turbulence may be affected by other factors such as the channel material and surface roughness. Besides the onset of large-scale normal fluid turbulence, increased frequency of particle-vortex reconnection may contribute to the larger velocity fluctuations of G3. Kivotides has shown through numerical simulation that in a dense vortex tangle, such reconnection events can induce velocity fluctuations of the particles comparable to their mean velocity~\cite{Kivotides2008b}. However, directly comparing the results is difficult since the simulations did not take into account the turbulent normal fluid. Finally, we note that the streamwise velocity standard deviation for G2, unlike the transverse, does seem to increase with heat flux. We will carry this observation forward into the following discussion.
	
	\section{Discussion}
	\label{sec:Discussion}
	
	There are a couple of potential explanations for the apparent anisotropy of G2 revealed in Fig.~\ref{fig:std}. This may be an artifact of the acceleration and deceleration of particles at the beginning and end, respectively, of G2 tracks when they break free from or become trapped on vortex lines. If this is indeed the correct physical interpretation, it implies that particles interact with vortices primarily through trapping, as opposed to wide angle scattering, as the latter would be associated with a significant acceleration, and thus velocity fluctuation, in the direction normal to the trajectory. It follows that the capture cross section significantly exceeds the wide angle scattering cross section. Numerical simulations that place a moving particle on a straight trajectory past a vortex line, with varying distance between the particle trajectory and vortex core, would reveal the trapping and scattering cross sections. Similar work has already been performed by Kivotides et al.~\cite{Kivotides2007,Kivotides2008} on the results of a direct collision between a moving particle and vortex line at relatively low temperature. Though expansion of this work to study ``near misses'' would be non-trivial, taking into account the vortex dynamics as well as the normal fluid drag force, it would provide important insight to the particle velocity statistics in thermal counterflow.
	
	A more feasible explanation is that since particles exhibiting G2 behavior move primarily under the influence of drag force from the normal fluid, their velocity is subject to local variations of the normal fluid velocity. As the normal fluid passes across the vortex tangle, wakes can form behind each individual vortex line due to mutual friction; within these wakes, the normal fluid velocity can vary significantly~\cite{VinenPrivate}. It makes sense that the same fluctuations do not appear in the transverse particle velocity since there is no mean flow in that direction.
	
	An additional point of interest is the dynamics of particles trapped in the vortex tangle. Our present work as well as that of Chagovets and Van Sciver~\cite{Chagovets2011} has shown that as the heat flux increases, G1 departs from $v_p\approx{v_s}$ behavior and transitions to roughly $v_n/2$. We do not have an explanation for this behavior, particularly, why G1 follows the same trend as G3. A similar departure from $v_p\approx{v_s}$ of trapped particles was first observed by Paoletti et al.~\cite{Paoletti2008JPS}, and has been reproduced in the numerical simulations of Mineda et al.~\cite{Mineda2013}, though neither presented the particle velocity in terms of $v_n$, and the simulation did not extend to values of $v_n$ very far beyond the transition point. It would be interesting to know whether a similar simulation, extended to heat currents farther beyond the transition, reveals the same evolution of mean G1 velocity with $v_n/2$ as the experiments, and why. 
	
	Furthermore, our experimental results suggest that the velocity fluctuations for G1 in both streamwise and transverse direction increase linearly with applied heat flux. To attempt an explanation we calculate the vortex line velocity fluctuation and compare the results with our G1 observations. Based on the localized induction approximation (LIA), the line velocity as a function of its local curvature $R$ can be written as~\cite{Vinen2002}:
	\begin{equation}
	\label{eq:lineVelocity}
	v_L=\frac{\kappa}{4\pi{R}}\ln\left(\frac{R}{\xi_0}\right)
	\end{equation}
	The line velocity fluctuation can then be obtained from $\left<v_L^2\right>^{1/2}$: 
	\begin{equation}
	\left<v_L^2\right>^{1/2}=\frac{\kappa}{4\pi}\left<\frac{1}{R^2}\ln^2\left(\frac{R}{\xi_0}\right)\right>^{1/2}
	\end{equation}
    Neglecting the slow variation of the natural logarithm with $L$, we make the substitution $R\approx\ell$ and remove the constant $\ln\left(\ell/\xi_0\right)$ from the average~\cite{GaoArxiv}. The remaining term $\left<1/R^2\right>^{1/2}$ can be replaced by $c_2L^{1/2}$, where $c_2$ is a temperature dependent parameter~\cite{Schwarz1988,Vinen2002,GaoArxiv}, and the line density can be written in terms of the normal fluid velocity as per~(\ref{eq:lineDensity}):
	\begin{equation}
	\left<\frac{1}{R^2}\right>^{1/2}\approx{c_2}\gamma\left(\frac{\rho}{\rho_s}v_n-v_0\right)
	\end{equation}
	The resulting expression for root mean square vortex line velocity fluctuation as a function of the normal fluid velocity is:
	\begin{equation}
	\label{eq:lineFluctuation}
	\begin{alignedat}{2}
	&\left<v_L^2\right>^{1/2}\approx\frac{\kappa{c_2}\gamma}{4\pi}\ln\left(\frac{\ell}{\xi_0}\right)\left(\frac{\rho}{\rho_s}v_n-v_0\right) &\mathrm{~~if~~} \frac{\rho}{\rho_s}v_n>v_0 \\
	&\left<v_L^2\right>^{1/2}=0 &\mathrm{~~if~~} \frac{\rho}{\rho_s}v_n\le{v_0}
	\end{alignedat}
	\end{equation}
	We note that this simple approach yields a linear relationship between $\left<v_L^2\right>^{1/2}$ and $v_n$, provided the counterflow velocity exceeds $v_0$. Using values for $c_2$~\cite{GaoArxiv} and $\gamma$~\cite{Gao2017} derived from the work of Gao et al., and an approximate value of $\ell\approx100~\mu$m, we find the proportionality constant to be 0.54. The solid blue line of Fig.~\ref{fig:std} represents~(\ref{eq:lineFluctuation}), with a small offset to adjust for environmental noise. It agrees reasonably well with the observed G1 velocity fluctuation, suggesting that the G1 particle velocity fluctuations are, to a good extent, caused by fluctuations of the vortex line velocity. However, it should be kept in mind that the particle vortex interaction is quite complicated, and depends also on such factors as the relative motion between the particles, vortices, and normal fluid, and deformation of the tangle due to the presence of particles. The same numerical simulation that predicts the mean velocity of trapped particles, suggested above, could produce more detailed information about the relationship between G1 and vortex tangle velocity fluctuations.
	
	Finally, in the future, we plan to study the structure and scaling laws of PDFs related to the particle trajectory geometry, instead of the kinematics. This approach to Lagrangian fluid dynamics has recently emerged in classical fluids, where curvature of the trajectories~\cite{HXu2007} or the relative angle of velocity vectors as a function of their temporal separation along the track~\cite{Burov2013} are used to characterize the fluid dynamics. Applications to simulations of classical turbulent flows have revealed power law scaling of the PDFs for trajectory angle~\cite{Bos2015}, curvature, and torsion~\cite{Bos2015,Bhatnagar2016} that characterize the turbulence. This approach has not yet been introduced to quantum turbulence, and may offer an opportunity for quantitative characterization of the vortex tangle dynamics in thermal counterflow~\cite{BarenghiPrivate}. This could of course be accomplished using experimental data as well as numerical simulations.

	\section{Conclusions}
	\label{sec:Conclusion}

	We have performed a systematic study of solidified particle motion in He II thermal counterflow using the PTV technique. For the first time in a single experiment, the driving heat flux extends from the low range, previously investigated by PTV, to the high range, previously investigated by PIV. Demonstrating that the streamwise velocity PDFs transform from a double peak structure, with one peak centered at $v_n$ and one near $v_n/2$, into a single peak centered near $v_n/2$, rectifies the previous experimental observations as well as predictions obtained through numerical simulations. 
	
	We have also devised a simple criteria to isolate the normal fluid and vortex tangle velocity statistics. We apply this separation criteria to show that G1 velocity measurements dominate the non-classical tail structure of the transverse velocity PDFs, while G2 velocity statistics exhibit more or less classical behavior. In order to better understand the observed behavior of each group, we hope that this work will stimulate a number of numerical simulations that further characterize particle motion in He II counterflow.
	
	\begin{acknowledgments}
		The authors wish to thank W.F. Vinen, C.F. Barenghi, and M. Tsubota for their valuable input on the discussion of our results. This work is supported by U.S. Department of Energy grant DE-FG02-96ER40952. It was conducted at the National High Magnetic Field Laboratory, which is supported by NSF DMR-1157490 and the State of Florida.
	\end{acknowledgments}
	

\begin{thebibliography}{66}%
\makeatletter
\providecommand \@ifxundefined [1]{%
 \@ifx{#1\undefined}
}%
\providecommand \@ifnum [1]{%
 \ifnum #1\expandafter \@firstoftwo
 \else \expandafter \@secondoftwo
 \fi
}%
\providecommand \@ifx [1]{%
 \ifx #1\expandafter \@firstoftwo
 \else \expandafter \@secondoftwo
 \fi
}%
\providecommand \natexlab [1]{#1}%
\providecommand \enquote  [1]{``#1''}%
\providecommand \bibnamefont  [1]{#1}%
\providecommand \bibfnamefont [1]{#1}%
\providecommand \citenamefont [1]{#1}%
\providecommand \href@noop [0]{\@secondoftwo}%
\providecommand \href [0]{\begingroup \@sanitize@url \@href}%
\providecommand \@href[1]{\@@startlink{#1}\@@href}%
\providecommand \@@href[1]{\endgroup#1\@@endlink}%
\providecommand \@sanitize@url [0]{\catcode `\\12\catcode `\$12\catcode
  `\&12\catcode `\#12\catcode `\^12\catcode `\_12\catcode `\%12\relax}%
\providecommand \@@startlink[1]{}%
\providecommand \@@endlink[0]{}%
\providecommand \url  [0]{\begingroup\@sanitize@url \@url }%
\providecommand \@url [1]{\endgroup\@href {#1}{\urlprefix }}%
\providecommand \urlprefix  [0]{URL }%
\providecommand \Eprint [0]{\href }%
\providecommand \doibase [0]{http://dx.doi.org/}%
\providecommand \selectlanguage [0]{\@gobble}%
\providecommand \bibinfo  [0]{\@secondoftwo}%
\providecommand \bibfield  [0]{\@secondoftwo}%
\providecommand \translation [1]{[#1]}%
\providecommand \BibitemOpen [0]{}%
\providecommand \bibitemStop [0]{}%
\providecommand \bibitemNoStop [0]{.\EOS\space}%
\providecommand \EOS [0]{\spacefactor3000\relax}%
\providecommand \BibitemShut  [1]{\csname bibitem#1\endcsname}%
\let\auto@bib@innerbib\@empty
\bibitem [{\citenamefont {Tisza}(1938)}]{Tisza1938}%
  \BibitemOpen
  \bibfield  {author} {\bibinfo {author} {\bibfnamefont {L.}~\bibnamefont
  {Tisza}},\ }\bibfield  {title} {\enquote {\bibinfo {title} {Transport
  phenomena in helium {II}},}\ }\href@noop {} {\bibfield  {journal} {\bibinfo
  {journal} {Nature}\ }\textbf {\bibinfo {volume} {141}},\ \bibinfo {pages}
  {913} (\bibinfo {year} {1938})}\BibitemShut {NoStop}%
\bibitem [{\citenamefont {Landau}(1941)}]{Landau1941}%
  \BibitemOpen
  \bibfield  {author} {\bibinfo {author} {\bibfnamefont {L.}~\bibnamefont
  {Landau}},\ }\bibfield  {title} {\enquote {\bibinfo {title} {Theory of the
  superfluidity of helium {II}},}\ }\href {\doibase 10.1103/PhysRev.60.356}
  {\bibfield  {journal} {\bibinfo  {journal} {Phys. Rev.}\ }\textbf {\bibinfo
  {volume} {60}},\ \bibinfo {pages} {356} (\bibinfo {year} {1941})}\BibitemShut
  {NoStop}%
\bibitem [{\citenamefont {Vinen}(2006)}]{Vinen2006}%
  \BibitemOpen
  \bibfield  {author} {\bibinfo {author} {\bibfnamefont {W.~F.}\ \bibnamefont
  {Vinen}},\ }\bibfield  {title} {\enquote {\bibinfo {title} {An introduction
  to quantum turbulence},}\ }\href {\doibase 10.1007/s10909-006-9240-6}
  {\bibfield  {journal} {\bibinfo  {journal} {J. Low Temp. Phys.}\ }\textbf
  {\bibinfo {volume} {145}},\ \bibinfo {pages} {7} (\bibinfo {year}
  {2006})}\BibitemShut {NoStop}%
\bibitem [{\citenamefont {Vinen}(1957{\natexlab{a}})}]{VinenII}%
  \BibitemOpen
  \bibfield  {author} {\bibinfo {author} {\bibfnamefont {W.~F.}\ \bibnamefont
  {Vinen}},\ }\bibfield  {title} {\enquote {\bibinfo {title} {Mutual friction
  in a heat current in liquid helium {II}. {II}. {E}xperiments on transient
  effects},}\ }\href {\doibase 10.1098/rspa.1957.0072} {\bibfield  {journal}
  {\bibinfo  {journal} {Proc. R. Soc. London, Ser. A}\ }\textbf {\bibinfo
  {volume} {240}},\ \bibinfo {pages} {128} (\bibinfo {year}
  {1957}{\natexlab{a}})}\BibitemShut {NoStop}%
\bibitem [{\citenamefont {Vinen}(1957{\natexlab{b}})}]{VinenIII}%
  \BibitemOpen
  \bibfield  {author} {\bibinfo {author} {\bibfnamefont {W.~F.}\ \bibnamefont
  {Vinen}},\ }\bibfield  {title} {\enquote {\bibinfo {title} {Mutual friction
  in a heat current in liquid helium {II}. {III}. {T}heory of the mutual
  friction},}\ }\href {\doibase 10.1098/rspa.1957.0191} {\bibfield  {journal}
  {\bibinfo  {journal} {Proc. R. Soc. London, Ser. A}\ }\textbf {\bibinfo
  {volume} {242}},\ \bibinfo {pages} {493} (\bibinfo {year}
  {1957}{\natexlab{b}})}\BibitemShut {NoStop}%
\bibitem [{\citenamefont {{Van Sciver}}(2012)}]{VanSciver2012}%
  \BibitemOpen
  \bibfield  {author} {\bibinfo {author} {\bibfnamefont {S.~W.}\ \bibnamefont
  {{Van Sciver}}},\ }\href@noop {} {\emph {\bibinfo {title} {Helium
  Cryogenics}}}\ (\bibinfo  {publisher} {Springer},\ \bibinfo {year}
  {2012})\BibitemShut {NoStop}%
\bibitem [{\citenamefont {Guo}\ \emph {et~al.}(2010)\citenamefont {Guo},
  \citenamefont {Cahn}, \citenamefont {Nikkel}, \citenamefont {Vinen},\ and\
  \citenamefont {McKinsey}}]{Guo2010}%
  \BibitemOpen
  \bibfield  {author} {\bibinfo {author} {\bibfnamefont {W.}~\bibnamefont
  {Guo}}, \bibinfo {author} {\bibfnamefont {S.~B.}\ \bibnamefont {Cahn}},
  \bibinfo {author} {\bibfnamefont {J.~A.}\ \bibnamefont {Nikkel}}, \bibinfo
  {author} {\bibfnamefont {W.~F.}\ \bibnamefont {Vinen}}, \ and\ \bibinfo
  {author} {\bibfnamefont {D.~N.}\ \bibnamefont {McKinsey}},\ }\bibfield
  {title} {\enquote {\bibinfo {title} {Visualization study of counterflow in
  superfluid $^{4}\mathrm{He}$ using metastable helium molecules},}\ }\href
  {\doibase 10.1103/PhysRevLett.105.045301} {\bibfield  {journal} {\bibinfo
  {journal} {Phys. Rev. Lett.}\ }\textbf {\bibinfo {volume} {105}},\ \bibinfo
  {pages} {045301} (\bibinfo {year} {2010})}\BibitemShut {NoStop}%
\bibitem [{\citenamefont {Marakov}\ \emph {et~al.}(2015)\citenamefont
  {Marakov}, \citenamefont {Gao}, \citenamefont {Guo}, \citenamefont
  {Van~Sciver}, \citenamefont {Ihas}, \citenamefont {McKinsey},\ and\
  \citenamefont {Vinen}}]{Marakov2015}%
  \BibitemOpen
  \bibfield  {author} {\bibinfo {author} {\bibfnamefont {A.}~\bibnamefont
  {Marakov}}, \bibinfo {author} {\bibfnamefont {J.}~\bibnamefont {Gao}},
  \bibinfo {author} {\bibfnamefont {W.}~\bibnamefont {Guo}}, \bibinfo {author}
  {\bibfnamefont {S.~W.}\ \bibnamefont {Van~Sciver}}, \bibinfo {author}
  {\bibfnamefont {G.~G.}\ \bibnamefont {Ihas}}, \bibinfo {author}
  {\bibfnamefont {D.~N.}\ \bibnamefont {McKinsey}}, \ and\ \bibinfo {author}
  {\bibfnamefont {W.~F.}\ \bibnamefont {Vinen}},\ }\bibfield  {title} {\enquote
  {\bibinfo {title} {Visualization of the normal-fluid turbulence in
  counterflowing superfluid $^{4}\mathrm{He}$},}\ }\href {\doibase
  10.1103/PhysRevB.91.094503} {\bibfield  {journal} {\bibinfo  {journal} {Phys.
  Rev. B}\ }\textbf {\bibinfo {volume} {91}},\ \bibinfo {pages} {094503}
  (\bibinfo {year} {2015})}\BibitemShut {NoStop}%
\bibitem [{\citenamefont {Chase}(1962)}]{Chase1962}%
  \BibitemOpen
  \bibfield  {author} {\bibinfo {author} {\bibfnamefont {C.~E.}\ \bibnamefont
  {Chase}},\ }\bibfield  {title} {\enquote {\bibinfo {title} {Thermal
  conduction in liquid helium {II}. {I}. {T}emperature dependence},}\ }\href
  {\doibase 10.1103/PhysRev.127.361} {\bibfield  {journal} {\bibinfo  {journal}
  {Phys. Rev.}\ }\textbf {\bibinfo {volume} {127}},\ \bibinfo {pages} {361}
  (\bibinfo {year} {1962})}\BibitemShut {NoStop}%
\bibitem [{\citenamefont {Dimotakis}\ and\ \citenamefont
  {Broadwell}(1973)}]{Dimotakis1973}%
  \BibitemOpen
  \bibfield  {author} {\bibinfo {author} {\bibfnamefont {P.~E.}\ \bibnamefont
  {Dimotakis}}\ and\ \bibinfo {author} {\bibfnamefont {J.~E.}\ \bibnamefont
  {Broadwell}},\ }\bibfield  {title} {\enquote {\bibinfo {title} {Local
  temperature measurements in supercritical counterflow in liquid helium
  {II}},}\ }\href {\doibase 10.1063/1.1694214} {\bibfield  {journal} {\bibinfo
  {journal} {Phys. Fluids}\ }\textbf {\bibinfo {volume} {16}},\ \bibinfo
  {pages} {1787} (\bibinfo {year} {1973})}\BibitemShut {NoStop}%
\bibitem [{\citenamefont {Childers}\ and\ \citenamefont
  {Tough}(1976)}]{Childers1976}%
  \BibitemOpen
  \bibfield  {author} {\bibinfo {author} {\bibfnamefont {R.~K.}\ \bibnamefont
  {Childers}}\ and\ \bibinfo {author} {\bibfnamefont {J.~T.}\ \bibnamefont
  {Tough}},\ }\bibfield  {title} {\enquote {\bibinfo {title} {Helium {II}
  thermal counterflow: Temperature- and pressure-difference data and analysis
  in terms of the {V}inen theory},}\ }\href {\doibase 10.1103/PhysRevB.13.1040}
  {\bibfield  {journal} {\bibinfo  {journal} {Phys. Rev. B}\ }\textbf {\bibinfo
  {volume} {13}},\ \bibinfo {pages} {1040} (\bibinfo {year}
  {1976})}\BibitemShut {NoStop}%
\bibitem [{\citenamefont {Martin}\ and\ \citenamefont
  {Tough}(1983)}]{Martin1983}%
  \BibitemOpen
  \bibfield  {author} {\bibinfo {author} {\bibfnamefont {K.~P.}\ \bibnamefont
  {Martin}}\ and\ \bibinfo {author} {\bibfnamefont {J.~T.}\ \bibnamefont
  {Tough}},\ }\bibfield  {title} {\enquote {\bibinfo {title} {Evolution of
  superfluid turbulence in thermal counterflow},}\ }\href {\doibase
  10.1103/PhysRevB.27.2788} {\bibfield  {journal} {\bibinfo  {journal} {Phys.
  Rev. B}\ }\textbf {\bibinfo {volume} {27}},\ \bibinfo {pages} {2788}
  (\bibinfo {year} {1983})}\BibitemShut {NoStop}%
\bibitem [{\citenamefont {Babuin}\ \emph {et~al.}(2012)\citenamefont {Babuin},
  \citenamefont {Stammeier}, \citenamefont {Varga}, \citenamefont {Rotter},\
  and\ \citenamefont {Skrbek}}]{Babuin2012}%
  \BibitemOpen
  \bibfield  {author} {\bibinfo {author} {\bibfnamefont {S.}~\bibnamefont
  {Babuin}}, \bibinfo {author} {\bibfnamefont {M.}~\bibnamefont {Stammeier}},
  \bibinfo {author} {\bibfnamefont {E.}~\bibnamefont {Varga}}, \bibinfo
  {author} {\bibfnamefont {M.}~\bibnamefont {Rotter}}, \ and\ \bibinfo {author}
  {\bibfnamefont {L.}~\bibnamefont {Skrbek}},\ }\bibfield  {title} {\enquote
  {\bibinfo {title} {Quantum turbulence of bellows-driven ${}^{4}${He}
  superflow: Steady state},}\ }\href {\doibase 10.1103/PhysRevB.86.134515}
  {\bibfield  {journal} {\bibinfo  {journal} {Phys. Rev. B}\ }\textbf {\bibinfo
  {volume} {86}},\ \bibinfo {pages} {134515} (\bibinfo {year}
  {2012})}\BibitemShut {NoStop}%
\bibitem [{\citenamefont {Gao}\ \emph {et~al.}(2017)\citenamefont {Gao},
  \citenamefont {Varga}, \citenamefont {Guo},\ and\ \citenamefont
  {Vinen}}]{Gao2017}%
  \BibitemOpen
  \bibfield  {author} {\bibinfo {author} {\bibfnamefont {J.}~\bibnamefont
  {Gao}}, \bibinfo {author} {\bibfnamefont {E.}~\bibnamefont {Varga}}, \bibinfo
  {author} {\bibfnamefont {W.}~\bibnamefont {Guo}}, \ and\ \bibinfo {author}
  {\bibfnamefont {W.~F.}\ \bibnamefont {Vinen}},\ }\bibfield  {title} {\enquote
  {\bibinfo {title} {Energy spectrum of thermal counterflow turbulence in
  superfluid helium-4},}\ }\href {\doibase 10.1103/PhysRevB.96.094511}
  {\bibfield  {journal} {\bibinfo  {journal} {Phys. Rev. B}\ }\textbf {\bibinfo
  {volume} {96}},\ \bibinfo {pages} {094511} (\bibinfo {year}
  {2017})}\BibitemShut {NoStop}%
\bibitem [{\citenamefont {Varga}\ \emph {et~al.}(2015)\citenamefont {Varga},
  \citenamefont {Babuin},\ and\ \citenamefont {Skrbek}}]{Varga2015}%
  \BibitemOpen
  \bibfield  {author} {\bibinfo {author} {\bibfnamefont {E.}~\bibnamefont
  {Varga}}, \bibinfo {author} {\bibfnamefont {S.}~\bibnamefont {Babuin}}, \
  and\ \bibinfo {author} {\bibfnamefont {L.}~\bibnamefont {Skrbek}},\
  }\bibfield  {title} {\enquote {\bibinfo {title} {Second-sound studies of
  coflow and counterflow of superfluid {4He} in channels},}\ }\href {\doibase
  10.1063/1.4921816} {\bibfield  {journal} {\bibinfo  {journal} {Phys. Fluids}\
  }\textbf {\bibinfo {volume} {27}},\ \bibinfo {pages} {065101} (\bibinfo
  {year} {2015})}\BibitemShut {NoStop}%
\bibitem [{\citenamefont {Guo}\ \emph {et~al.}(2014)\citenamefont {Guo},
  \citenamefont {La~Mantia}, \citenamefont {Lathrop},\ and\ \citenamefont
  {Van~Sciver}}]{Guo2014}%
  \BibitemOpen
  \bibfield  {author} {\bibinfo {author} {\bibfnamefont {W.}~\bibnamefont
  {Guo}}, \bibinfo {author} {\bibfnamefont {M.}~\bibnamefont {La~Mantia}},
  \bibinfo {author} {\bibfnamefont {D.~P.}\ \bibnamefont {Lathrop}}, \ and\
  \bibinfo {author} {\bibfnamefont {S.~W.}\ \bibnamefont {Van~Sciver}},\
  }\bibfield  {title} {\enquote {\bibinfo {title} {Visualization of two-fluid
  flows of superfluid helium-4},}\ }\href {\doibase 10.1073/pnas.1312546111}
  {\bibfield  {journal} {\bibinfo  {journal} {Proc. Natl. Acad. Sci. USA}\
  }\textbf {\bibinfo {volume} {111}},\ \bibinfo {pages} {4653} (\bibinfo {year}
  {2014})}\BibitemShut {NoStop}%
\bibitem [{\citenamefont {Raffel}\ \emph {et~al.}(1998)\citenamefont {Raffel},
  \citenamefont {Willert},\ and\ \citenamefont {Kompenhans}}]{Raffel1998}%
  \BibitemOpen
  \bibfield  {author} {\bibinfo {author} {\bibfnamefont {M.}~\bibnamefont
  {Raffel}}, \bibinfo {author} {\bibfnamefont {C.}~\bibnamefont {Willert}}, \
  and\ \bibinfo {author} {\bibfnamefont {J.}~\bibnamefont {Kompenhans}},\
  }\href@noop {} {\emph {\bibinfo {title} {Particle Image Velocimetry}}}\
  (\bibinfo  {publisher} {Springer},\ \bibinfo {year} {1998})\BibitemShut
  {NoStop}%
\bibitem [{\citenamefont {Sergeev}\ and\ \citenamefont
  {Barenghi}(2009)}]{Sergeev2009}%
  \BibitemOpen
  \bibfield  {author} {\bibinfo {author} {\bibfnamefont {Y.~A.}\ \bibnamefont
  {Sergeev}}\ and\ \bibinfo {author} {\bibfnamefont {C.~F.}\ \bibnamefont
  {Barenghi}},\ }\bibfield  {title} {\enquote {\bibinfo {title}
  {Particles-vortex interactions and flow visualization in {4He}},}\ }\href
  {\doibase 10.1007/s10909-009-9994-8} {\bibfield  {journal} {\bibinfo
  {journal} {J. Low Temp. Phys.}\ }\textbf {\bibinfo {volume} {157}},\ \bibinfo
  {pages} {429} (\bibinfo {year} {2009})}\BibitemShut {NoStop}%
\bibitem [{\citenamefont {Parks}\ and\ \citenamefont
  {Donnelly}(1966)}]{Parks1966}%
  \BibitemOpen
  \bibfield  {author} {\bibinfo {author} {\bibfnamefont {P.~E.}\ \bibnamefont
  {Parks}}\ and\ \bibinfo {author} {\bibfnamefont {R.~J.}\ \bibnamefont
  {Donnelly}},\ }\bibfield  {title} {\enquote {\bibinfo {title} {Radii of
  positive and negative ions in helium {II}},}\ }\href {\doibase
  10.1103/PhysRevLett.16.45} {\bibfield  {journal} {\bibinfo  {journal} {Phys.
  Rev. Lett.}\ }\textbf {\bibinfo {volume} {16}},\ \bibinfo {pages} {45}
  (\bibinfo {year} {1966})}\BibitemShut {NoStop}%
\bibitem [{\citenamefont {Bewley}\ \emph {et~al.}(2006)\citenamefont {Bewley},
  \citenamefont {Lathrop},\ and\ \citenamefont {Sreenivasan}}]{Bewley2006}%
  \BibitemOpen
  \bibfield  {author} {\bibinfo {author} {\bibfnamefont {G.~P.}\ \bibnamefont
  {Bewley}}, \bibinfo {author} {\bibfnamefont {D.~P.}\ \bibnamefont {Lathrop}},
  \ and\ \bibinfo {author} {\bibfnamefont {K.~R.}\ \bibnamefont
  {Sreenivasan}},\ }\bibfield  {title} {\enquote {\bibinfo {title} {Superfluid
  helium: Visualization of quantized vortices},}\ }\href {\doibase
  10.1038/441588a} {\bibfield  {journal} {\bibinfo  {journal} {Nature}\
  }\textbf {\bibinfo {volume} {441}},\ \bibinfo {pages} {588} (\bibinfo {year}
  {2006})}\BibitemShut {NoStop}%
\bibitem [{\citenamefont {Kivotides}(2008{\natexlab{a}})}]{Kivotides2008c}%
  \BibitemOpen
  \bibfield  {author} {\bibinfo {author} {\bibfnamefont {D.}~\bibnamefont
  {Kivotides}},\ }\bibfield  {title} {\enquote {\bibinfo {title} {Normal-fluid
  velocity measurement and superfluid vortex detection in thermal counterflow
  turbulence},}\ }\href {\doibase 10.1103/PhysRevB.78.224501} {\bibfield
  {journal} {\bibinfo  {journal} {Phys. Rev. B}\ }\textbf {\bibinfo {volume}
  {78}},\ \bibinfo {pages} {224501} (\bibinfo {year}
  {2008}{\natexlab{a}})}\BibitemShut {NoStop}%
\bibitem [{\citenamefont {Mineda}\ \emph {et~al.}(2013)\citenamefont {Mineda},
  \citenamefont {Tsubota}, \citenamefont {Sergeev}, \citenamefont {Barenghi},\
  and\ \citenamefont {Vinen}}]{Mineda2013}%
  \BibitemOpen
  \bibfield  {author} {\bibinfo {author} {\bibfnamefont {Y.}~\bibnamefont
  {Mineda}}, \bibinfo {author} {\bibfnamefont {M.}~\bibnamefont {Tsubota}},
  \bibinfo {author} {\bibfnamefont {Y.~A.}\ \bibnamefont {Sergeev}}, \bibinfo
  {author} {\bibfnamefont {C.~F.}\ \bibnamefont {Barenghi}}, \ and\ \bibinfo
  {author} {\bibfnamefont {W.~F.}\ \bibnamefont {Vinen}},\ }\bibfield  {title}
  {\enquote {\bibinfo {title} {Velocity distributions of tracer particles in
  thermal counterflow in superfluid ${}^{4}${He}},}\ }\href {\doibase
  10.1103/PhysRevB.87.174508} {\bibfield  {journal} {\bibinfo  {journal} {Phys.
  Rev. B}\ }\textbf {\bibinfo {volume} {87}},\ \bibinfo {pages} {174508}
  (\bibinfo {year} {2013})}\BibitemShut {NoStop}%
\bibitem [{\citenamefont {Zhang}\ and\ \citenamefont
  {Van~Sciver}(2005)}]{Zhang2005}%
  \BibitemOpen
  \bibfield  {author} {\bibinfo {author} {\bibfnamefont {T.}~\bibnamefont
  {Zhang}}\ and\ \bibinfo {author} {\bibfnamefont {S.~W.}\ \bibnamefont
  {Van~Sciver}},\ }\bibfield  {title} {\enquote {\bibinfo {title} {The motion
  of micron-sized particles in {He II} counterflow as observed by the {PIV}
  technique},}\ }\href {\doibase 10.1007/s10909-005-2316-x} {\bibfield
  {journal} {\bibinfo  {journal} {J. Low Temp. Phys.}\ }\textbf {\bibinfo
  {volume} {138}},\ \bibinfo {pages} {865} (\bibinfo {year}
  {2005})}\BibitemShut {NoStop}%
\bibitem [{\citenamefont {Sergeev}\ \emph {et~al.}(2006)\citenamefont
  {Sergeev}, \citenamefont {Barenghi},\ and\ \citenamefont
  {Kivotides}}]{Sergeev2006}%
  \BibitemOpen
  \bibfield  {author} {\bibinfo {author} {\bibfnamefont {Y.~A.}\ \bibnamefont
  {Sergeev}}, \bibinfo {author} {\bibfnamefont {C.~F.}\ \bibnamefont
  {Barenghi}}, \ and\ \bibinfo {author} {\bibfnamefont {D.}~\bibnamefont
  {Kivotides}},\ }\bibfield  {title} {\enquote {\bibinfo {title} {Motion of
  micron-size particles in turbulent helium {II}},}\ }\href {\doibase
  10.1103/PhysRevB.74.184506} {\bibfield  {journal} {\bibinfo  {journal} {Phys.
  Rev. B}\ }\textbf {\bibinfo {volume} {74}},\ \bibinfo {pages} {184506}
  (\bibinfo {year} {2006})}\BibitemShut {NoStop}%
\bibitem [{\citenamefont {Paoletti}\ \emph
  {et~al.}(2008{\natexlab{a}})\citenamefont {Paoletti}, \citenamefont
  {Fiorito}, \citenamefont {Sreenivasan},\ and\ \citenamefont
  {Lathrop}}]{Paoletti2008JPS}%
  \BibitemOpen
  \bibfield  {author} {\bibinfo {author} {\bibfnamefont {M.~S.}\ \bibnamefont
  {Paoletti}}, \bibinfo {author} {\bibfnamefont {R.~B.}\ \bibnamefont
  {Fiorito}}, \bibinfo {author} {\bibfnamefont {K.~R.}\ \bibnamefont
  {Sreenivasan}}, \ and\ \bibinfo {author} {\bibfnamefont {D.~P.}\ \bibnamefont
  {Lathrop}},\ }\bibfield  {title} {\enquote {\bibinfo {title} {Visualization
  of superfluid helium flow},}\ }\href {\doibase 10.1143/JPSJ.77.111007}
  {\bibfield  {journal} {\bibinfo  {journal} {J. Phys. Soc. Jpn.}\ }\textbf
  {\bibinfo {volume} {77}},\ \bibinfo {pages} {111007} (\bibinfo {year}
  {2008}{\natexlab{a}})}\BibitemShut {NoStop}%
\bibitem [{\citenamefont {Chagovets}\ and\ \citenamefont {{Van
  Sciver}}(2011)}]{Chagovets2011}%
  \BibitemOpen
  \bibfield  {author} {\bibinfo {author} {\bibfnamefont {T.~V.}\ \bibnamefont
  {Chagovets}}\ and\ \bibinfo {author} {\bibfnamefont {S.~W.}\ \bibnamefont
  {{Van Sciver}}},\ }\bibfield  {title} {\enquote {\bibinfo {title} {A study of
  thermal counterflow using particle tracking velocimetry},}\ }\href {\doibase
  10.1063/1.3657084} {\bibfield  {journal} {\bibinfo  {journal} {Phys. Fluids}\
  }\textbf {\bibinfo {volume} {23}},\ \bibinfo {pages} {107102} (\bibinfo
  {year} {2011})}\BibitemShut {NoStop}%
\bibitem [{\citenamefont {{La Mantia}}(2016)}]{LaMantia2016}%
  \BibitemOpen
  \bibfield  {author} {\bibinfo {author} {\bibfnamefont {M.}~\bibnamefont {{La
  Mantia}}},\ }\bibfield  {title} {\enquote {\bibinfo {title} {Particle
  trajectories in thermal counterflow of superfluid helium in a wide channel of
  square cross section},}\ }\href {\doibase 10.1063/1.4940980} {\bibfield
  {journal} {\bibinfo  {journal} {Phys. Fluids}\ }\textbf {\bibinfo {volume}
  {28}},\ \bibinfo {pages} {024102} (\bibinfo {year} {2016})}\BibitemShut
  {NoStop}%
\bibitem [{\citenamefont {Kubo}\ and\ \citenamefont {Tsuji}(2017)}]{Kubo2017}%
  \BibitemOpen
  \bibfield  {author} {\bibinfo {author} {\bibfnamefont {W.}~\bibnamefont
  {Kubo}}\ and\ \bibinfo {author} {\bibfnamefont {Y.}~\bibnamefont {Tsuji}},\
  }\bibfield  {title} {\enquote {\bibinfo {title} {Lagrangian trajectory of
  small particles in superfluid {He II}},}\ }\href {\doibase
  10.1007/s10909-017-1764-4} {\bibfield  {journal} {\bibinfo  {journal} {J. Low
  Temp. Phys.}\ }\textbf {\bibinfo {volume} {187}},\ \bibinfo {pages} {611}
  (\bibinfo {year} {2017})}\BibitemShut {NoStop}%
\bibitem [{\citenamefont {Kivotides}(2008{\natexlab{b}})}]{Kivotides2008b}%
  \BibitemOpen
  \bibfield  {author} {\bibinfo {author} {\bibfnamefont {D.}~\bibnamefont
  {Kivotides}},\ }\bibfield  {title} {\enquote {\bibinfo {title} {Motion of a
  spherical solid particle in thermal counterflow turbulence},}\ }\href
  {\doibase 10.1103/PhysRevB.77.174508} {\bibfield  {journal} {\bibinfo
  {journal} {Phys. Rev. B}\ }\textbf {\bibinfo {volume} {77}},\ \bibinfo
  {pages} {174508} (\bibinfo {year} {2008}{\natexlab{b}})}\BibitemShut
  {NoStop}%
\bibitem [{\citenamefont {Mantia}\ and\ \citenamefont
  {Skrbek}(2014)}]{LaMantia2014a}%
  \BibitemOpen
  \bibfield  {author} {\bibinfo {author} {\bibfnamefont {M.~La}\ \bibnamefont
  {Mantia}}\ and\ \bibinfo {author} {\bibfnamefont {L.}~\bibnamefont
  {Skrbek}},\ }\bibfield  {title} {\enquote {\bibinfo {title} {Quantum, or
  classical turbulence?}}\ }\href@noop {} {\bibfield  {journal} {\bibinfo
  {journal} {Europhys. Lett.}\ }\textbf {\bibinfo {volume} {105}},\ \bibinfo
  {pages} {46002} (\bibinfo {year} {2014})}\BibitemShut {NoStop}%
\bibitem [{\citenamefont {Paoletti}\ \emph
  {et~al.}(2008{\natexlab{b}})\citenamefont {Paoletti}, \citenamefont {Fisher},
  \citenamefont {Sreenivasan},\ and\ \citenamefont {Lathrop}}]{Paoletti2008}%
  \BibitemOpen
  \bibfield  {author} {\bibinfo {author} {\bibfnamefont {M.~S.}\ \bibnamefont
  {Paoletti}}, \bibinfo {author} {\bibfnamefont {M.~E.}\ \bibnamefont
  {Fisher}}, \bibinfo {author} {\bibfnamefont {K.~R.}\ \bibnamefont
  {Sreenivasan}}, \ and\ \bibinfo {author} {\bibfnamefont {D.~P.}\ \bibnamefont
  {Lathrop}},\ }\bibfield  {title} {\enquote {\bibinfo {title} {Velocity
  statistics distinguish quantum turbulence from classical turbulence},}\
  }\href {\doibase 10.1103/PhysRevLett.101.154501} {\bibfield  {journal}
  {\bibinfo  {journal} {Phys. Rev. Lett.}\ }\textbf {\bibinfo {volume} {101}},\
  \bibinfo {pages} {154501} (\bibinfo {year} {2008}{\natexlab{b}})}\BibitemShut
  {NoStop}%
\bibitem [{\citenamefont {La~Mantia}\ and\ \citenamefont
  {Skrbek}(2014)}]{LaMantia2014b}%
  \BibitemOpen
  \bibfield  {author} {\bibinfo {author} {\bibfnamefont {M.}~\bibnamefont
  {La~Mantia}}\ and\ \bibinfo {author} {\bibfnamefont {L.}~\bibnamefont
  {Skrbek}},\ }\bibfield  {title} {\enquote {\bibinfo {title} {Quantum
  turbulence visualized by particle dynamics},}\ }\href {\doibase
  10.1103/PhysRevB.90.014519} {\bibfield  {journal} {\bibinfo  {journal} {Phys.
  Rev. B}\ }\textbf {\bibinfo {volume} {90}},\ \bibinfo {pages} {014519}
  (\bibinfo {year} {2014})}\BibitemShut {NoStop}%
\bibitem [{\citenamefont {Barenghi}\ \emph {et~al.}(2007)\citenamefont
  {Barenghi}, \citenamefont {Kivotides},\ and\ \citenamefont
  {Sergeev}}]{Barenghi2007}%
  \BibitemOpen
  \bibfield  {author} {\bibinfo {author} {\bibfnamefont {C.~F.}\ \bibnamefont
  {Barenghi}}, \bibinfo {author} {\bibfnamefont {D.}~\bibnamefont {Kivotides}},
  \ and\ \bibinfo {author} {\bibfnamefont {Y.~A.}\ \bibnamefont {Sergeev}},\
  }\bibfield  {title} {\enquote {\bibinfo {title} {Close approach of a
  spherical particle and a quantised vortex in helium {II}},}\ }\href {\doibase
  10.1007/s10909-007-9387-9} {\bibfield  {journal} {\bibinfo  {journal} {J. Low
  Temp. Phys.}\ }\textbf {\bibinfo {volume} {148}},\ \bibinfo {pages} {293}
  (\bibinfo {year} {2007})}\BibitemShut {NoStop}%
\bibitem [{\citenamefont {Kivotides}\ \emph {et~al.}(2008)\citenamefont
  {Kivotides}, \citenamefont {Barenghi},\ and\ \citenamefont
  {Sergeev}}]{Kivotides2008}%
  \BibitemOpen
  \bibfield  {author} {\bibinfo {author} {\bibfnamefont {D.}~\bibnamefont
  {Kivotides}}, \bibinfo {author} {\bibfnamefont {C.~F.}\ \bibnamefont
  {Barenghi}}, \ and\ \bibinfo {author} {\bibfnamefont {Y.~A.}\ \bibnamefont
  {Sergeev}},\ }\bibfield  {title} {\enquote {\bibinfo {title} {Interactions
  between particles and quantized vortices in superfluid helium},}\ }\href
  {\doibase 10.1103/PhysRevB.77.014527} {\bibfield  {journal} {\bibinfo
  {journal} {Phys. Rev. B}\ }\textbf {\bibinfo {volume} {77}},\ \bibinfo
  {pages} {014527} (\bibinfo {year} {2008})}\BibitemShut {NoStop}%
\bibitem [{\citenamefont {Zhang}\ and\ \citenamefont {{Van
  Sciver}}(2005)}]{Zhang2005b}%
  \BibitemOpen
  \bibfield  {author} {\bibinfo {author} {\bibfnamefont {T.}~\bibnamefont
  {Zhang}}\ and\ \bibinfo {author} {\bibfnamefont {S.~W.}\ \bibnamefont {{Van
  Sciver}}},\ }\bibfield  {title} {\enquote {\bibinfo {title} {Large-scale
  turbulent flow around a cylinder in counterflow superfluid
  \textsuperscript{4}{He} {(He(II))}},}\ }\href@noop {} {\bibfield  {journal}
  {\bibinfo  {journal} {Nature Phys.}\ }\textbf {\bibinfo {volume} {1}},\
  \bibinfo {pages} {36} (\bibinfo {year} {2005})}\BibitemShut {NoStop}%
\bibitem [{\citenamefont {Chagovets}\ and\ \citenamefont
  {Sciver}(2013)}]{Chagovets2013}%
  \BibitemOpen
  \bibfield  {author} {\bibinfo {author} {\bibfnamefont {T.~V.}\ \bibnamefont
  {Chagovets}}\ and\ \bibinfo {author} {\bibfnamefont {S.~W.~Van}\ \bibnamefont
  {Sciver}},\ }\bibfield  {title} {\enquote {\bibinfo {title} {Visualization of
  {He II} counterflow around a cylinder},}\ }\href {\doibase 10.1063/1.4824004}
  {\bibfield  {journal} {\bibinfo  {journal} {Phys. Fluids}\ }\textbf {\bibinfo
  {volume} {25}},\ \bibinfo {pages} {105104} (\bibinfo {year}
  {2013})}\BibitemShut {NoStop}%
\bibitem [{\citenamefont {Duda}\ \emph {et~al.}(2014)\citenamefont {Duda},
  \citenamefont {La~Mantia}, \citenamefont {Rotter},\ and\ \citenamefont
  {Skrbek}}]{Duda2014}%
  \BibitemOpen
  \bibfield  {author} {\bibinfo {author} {\bibfnamefont {D.}~\bibnamefont
  {Duda}}, \bibinfo {author} {\bibfnamefont {M.}~\bibnamefont {La~Mantia}},
  \bibinfo {author} {\bibfnamefont {M.}~\bibnamefont {Rotter}}, \ and\ \bibinfo
  {author} {\bibfnamefont {L.}~\bibnamefont {Skrbek}},\ }\bibfield  {title}
  {\enquote {\bibinfo {title} {On the visualization of thermal counterflow of
  {He II} past a circular cylinder},}\ }\href {\doibase
  10.1007/s10909-013-0961-z} {\bibfield  {journal} {\bibinfo  {journal} {J. Low
  Temp. Phys.}\ }\textbf {\bibinfo {volume} {175}},\ \bibinfo {pages} {331}
  (\bibinfo {year} {2014})}\BibitemShut {NoStop}%
\bibitem [{\citenamefont {Chagovets}\ and\ \citenamefont
  {Sciver}(2015)}]{Chagovets2015}%
  \BibitemOpen
  \bibfield  {author} {\bibinfo {author} {\bibfnamefont {T.~V.}\ \bibnamefont
  {Chagovets}}\ and\ \bibinfo {author} {\bibfnamefont {S.~W.~Van}\ \bibnamefont
  {Sciver}},\ }\bibfield  {title} {\enquote {\bibinfo {title} {Visualization of
  {He II} forced flow around a cylinder},}\ }\href {\doibase 10.1063/1.4919341}
  {\bibfield  {journal} {\bibinfo  {journal} {Phys. Fluids}\ }\textbf {\bibinfo
  {volume} {27}},\ \bibinfo {pages} {045111} (\bibinfo {year}
  {2015})}\BibitemShut {NoStop}%
\bibitem [{\citenamefont {Xu}\ and\ \citenamefont {Sciver}(2007)}]{Xu2007}%
  \BibitemOpen
  \bibfield  {author} {\bibinfo {author} {\bibfnamefont {T.}~\bibnamefont
  {Xu}}\ and\ \bibinfo {author} {\bibfnamefont {S.~W.~Van}\ \bibnamefont
  {Sciver}},\ }\bibfield  {title} {\enquote {\bibinfo {title} {Particle image
  velocimetry measurements of the velocity profile in {HeII} forced flow},}\
  }\href {\doibase 10.1063/1.2756577} {\bibfield  {journal} {\bibinfo
  {journal} {Phys. Fluids}\ }\textbf {\bibinfo {volume} {19}},\ \bibinfo
  {pages} {071703} (\bibinfo {year} {2007})}\BibitemShut {NoStop}%
\bibitem [{\citenamefont {Bewley}\ \emph {et~al.}(2008)\citenamefont {Bewley},
  \citenamefont {Paoletti}, \citenamefont {Sreenivasan},\ and\ \citenamefont
  {Lathrop}}]{Bewley2008}%
  \BibitemOpen
  \bibfield  {author} {\bibinfo {author} {\bibfnamefont {G.~P.}\ \bibnamefont
  {Bewley}}, \bibinfo {author} {\bibfnamefont {M.~S.}\ \bibnamefont
  {Paoletti}}, \bibinfo {author} {\bibfnamefont {K.~R.}\ \bibnamefont
  {Sreenivasan}}, \ and\ \bibinfo {author} {\bibfnamefont {D.~P.}\ \bibnamefont
  {Lathrop}},\ }\bibfield  {title} {\enquote {\bibinfo {title}
  {Characterization of reconnecting vortices in superfluid helium},}\ }\href
  {\doibase 10.1073/pnas.0806002105} {\bibfield  {journal} {\bibinfo  {journal}
  {Proc. Natl. Acad. Sci. USA}\ }\textbf {\bibinfo {volume} {105}},\ \bibinfo
  {pages} {13707} (\bibinfo {year} {2008})}\BibitemShut {NoStop}%
\bibitem [{\citenamefont {Paoletti}\ \emph {et~al.}(2010)\citenamefont
  {Paoletti}, \citenamefont {Fisher},\ and\ \citenamefont
  {Lathrop}}]{Paoletti2010}%
  \BibitemOpen
  \bibfield  {author} {\bibinfo {author} {\bibfnamefont {M.~S.}\ \bibnamefont
  {Paoletti}}, \bibinfo {author} {\bibfnamefont {M.~E.}\ \bibnamefont
  {Fisher}}, \ and\ \bibinfo {author} {\bibfnamefont {D.~P.}\ \bibnamefont
  {Lathrop}},\ }\bibfield  {title} {\enquote {\bibinfo {title} {Reconnection
  dynamics for quantized vortices},}\ }\href {\doibase
  https://doi.org/10.1016/j.physd.2009.03.006} {\bibfield  {journal} {\bibinfo
  {journal} {Physica D}\ }\textbf {\bibinfo {volume} {239}},\ \bibinfo {pages}
  {1367} (\bibinfo {year} {2010})}\BibitemShut {NoStop}%
\bibitem [{\citenamefont {Fonda}\ \emph {et~al.}(2014)\citenamefont {Fonda},
  \citenamefont {Meichle}, \citenamefont {Ouellette}, \citenamefont {Hormoz},\
  and\ \citenamefont {Lathrop}}]{Fonda2014}%
  \BibitemOpen
  \bibfield  {author} {\bibinfo {author} {\bibfnamefont {E.}~\bibnamefont
  {Fonda}}, \bibinfo {author} {\bibfnamefont {D.~P.}\ \bibnamefont {Meichle}},
  \bibinfo {author} {\bibfnamefont {N.~T.}\ \bibnamefont {Ouellette}}, \bibinfo
  {author} {\bibfnamefont {S.}~\bibnamefont {Hormoz}}, \ and\ \bibinfo {author}
  {\bibfnamefont {D.~P.}\ \bibnamefont {Lathrop}},\ }\bibfield  {title}
  {\enquote {\bibinfo {title} {Direct observation of kelvin waves excited by
  quantized vortex reconnection},}\ }\href {\doibase 10.1073/pnas.1312536110}
  {\bibfield  {journal} {\bibinfo  {journal} {Proc. Natl. Acad. Sci.}\ }\textbf
  {\bibinfo {volume} {111}},\ \bibinfo {pages} {4707} (\bibinfo {year}
  {2014})}\BibitemShut {NoStop}%
\bibitem [{\citenamefont {\ifmmode \check{S}\else
  \v{S}\fi{}van\ifmmode~\check{c}\else \v{c}\fi{}ara}\ and\ \citenamefont {{La
  Mantia}}(2017)}]{Svancara2017}%
  \BibitemOpen
  \bibfield  {author} {\bibinfo {author} {\bibfnamefont {P.}~\bibnamefont
  {\ifmmode \check{S}\else \v{S}\fi{}van\ifmmode~\check{c}\else
  \v{c}\fi{}ara}}\ and\ \bibinfo {author} {\bibfnamefont {M.}~\bibnamefont {{La
  Mantia}}},\ }\bibfield  {title} {\enquote {\bibinfo {title} {Flows of liquid
  {4He} due to oscillating grids},}\ }\href {\doibase 10.1017/jfm.2017.703}
  {\bibfield  {journal} {\bibinfo  {journal} {J. Fluid Mech.}\ }\textbf
  {\bibinfo {volume} {832}},\ \bibinfo {pages} {578} (\bibinfo {year}
  {2017})}\BibitemShut {NoStop}%
\bibitem [{\citenamefont {Mastracci}\ and\ \citenamefont
  {Guo}(2018)}]{Mastracci2018}%
  \BibitemOpen
  \bibfield  {author} {\bibinfo {author} {\bibfnamefont {B.}~\bibnamefont
  {Mastracci}}\ and\ \bibinfo {author} {\bibfnamefont {W.}~\bibnamefont
  {Guo}},\ }\bibfield  {title} {\enquote {\bibinfo {title} {An apparatus for
  generation and quantitative measurement of homogeneous isotropic turbulence
  in {He II}},}\ }\href {\doibase 10.1063/1.4997735} {\bibfield  {journal}
  {\bibinfo  {journal} {Rev. Sci. Instrum.}\ }\textbf {\bibinfo {volume}
  {89}},\ \bibinfo {pages} {015107} (\bibinfo {year} {2018})}\BibitemShut
  {NoStop}%
\bibitem [{\citenamefont {Gao}\ \emph {et~al.}(2015)\citenamefont {Gao},
  \citenamefont {Marakov}, \citenamefont {Guo}, \citenamefont {Pawlowski},
  \citenamefont {Sciver}, \citenamefont {Ihas}, \citenamefont {McKinsey},\ and\
  \citenamefont {Vinen}}]{Gao2015}%
  \BibitemOpen
  \bibfield  {author} {\bibinfo {author} {\bibfnamefont {J.}~\bibnamefont
  {Gao}}, \bibinfo {author} {\bibfnamefont {A.}~\bibnamefont {Marakov}},
  \bibinfo {author} {\bibfnamefont {W.}~\bibnamefont {Guo}}, \bibinfo {author}
  {\bibfnamefont {B.~T.}\ \bibnamefont {Pawlowski}}, \bibinfo {author}
  {\bibfnamefont {S.~W.~Van}\ \bibnamefont {Sciver}}, \bibinfo {author}
  {\bibfnamefont {G.~G.}\ \bibnamefont {Ihas}}, \bibinfo {author}
  {\bibfnamefont {D.~N.}\ \bibnamefont {McKinsey}}, \ and\ \bibinfo {author}
  {\bibfnamefont {W.~F.}\ \bibnamefont {Vinen}},\ }\bibfield  {title} {\enquote
  {\bibinfo {title} {Producing and imaging a thin line of {He2∗} molecular
  tracers in helium-4},}\ }\href {\doibase 10.1063/1.4930147} {\bibfield
  {journal} {\bibinfo  {journal} {Rev. Sci. Instrum.}\ }\textbf {\bibinfo
  {volume} {86}},\ \bibinfo {pages} {093904} (\bibinfo {year}
  {2015})}\BibitemShut {NoStop}%
\bibitem [{\citenamefont {Gao}\ \emph {et~al.}(2016{\natexlab{a}})\citenamefont
  {Gao}, \citenamefont {Guo},\ and\ \citenamefont {Vinen}}]{Gao2016}%
  \BibitemOpen
  \bibfield  {author} {\bibinfo {author} {\bibfnamefont {J.}~\bibnamefont
  {Gao}}, \bibinfo {author} {\bibfnamefont {W.}~\bibnamefont {Guo}}, \ and\
  \bibinfo {author} {\bibfnamefont {W.~F.}\ \bibnamefont {Vinen}},\ }\bibfield
  {title} {\enquote {\bibinfo {title} {Determination of the effective kinematic
  viscosity for the decay of quasiclassical turbulence in superfluid
  $^{4}\mathrm{He}$},}\ }\href {\doibase 10.1103/PhysRevB.94.094502} {\bibfield
   {journal} {\bibinfo  {journal} {Phys. Rev. B}\ }\textbf {\bibinfo {volume}
  {94}},\ \bibinfo {pages} {094502} (\bibinfo {year}
  {2016}{\natexlab{a}})}\BibitemShut {NoStop}%
\bibitem [{\citenamefont {Fonda}\ \emph {et~al.}(2016)\citenamefont {Fonda},
  \citenamefont {Sreenivasan},\ and\ \citenamefont {Lathrop}}]{Fonda2016}%
  \BibitemOpen
  \bibfield  {author} {\bibinfo {author} {\bibfnamefont {E.}~\bibnamefont
  {Fonda}}, \bibinfo {author} {\bibfnamefont {K.~R.}\ \bibnamefont
  {Sreenivasan}}, \ and\ \bibinfo {author} {\bibfnamefont {D.~P.}\ \bibnamefont
  {Lathrop}},\ }\bibfield  {title} {\enquote {\bibinfo {title} {Sub-micron
  solid air tracers for quantum vortices and liquid helium flows},}\ }\href
  {\doibase 10.1063/1.4941337} {\bibfield  {journal} {\bibinfo  {journal} {Rev.
  Sci. Instrum.}\ }\textbf {\bibinfo {volume} {87}},\ \bibinfo {pages} {025106}
  (\bibinfo {year} {2016})}\BibitemShut {NoStop}%
\bibitem [{\citenamefont {Soulaine}\ \emph {et~al.}(2017)\citenamefont
  {Soulaine}, \citenamefont {Quintard}, \citenamefont {Baudouy},\ and\
  \citenamefont {Van~Weelderen}}]{Soulaine2017}%
  \BibitemOpen
  \bibfield  {author} {\bibinfo {author} {\bibfnamefont {C.}~\bibnamefont
  {Soulaine}}, \bibinfo {author} {\bibfnamefont {M.}~\bibnamefont {Quintard}},
  \bibinfo {author} {\bibfnamefont {B.}~\bibnamefont {Baudouy}}, \ and\
  \bibinfo {author} {\bibfnamefont {R.}~\bibnamefont {Van~Weelderen}},\
  }\bibfield  {title} {\enquote {\bibinfo {title} {Numerical investigation of
  thermal counterflow of {He II} past cylinders},}\ }\href {\doibase
  10.1103/PhysRevLett.118.074506} {\bibfield  {journal} {\bibinfo  {journal}
  {Phys. Rev. Lett.}\ }\textbf {\bibinfo {volume} {118}},\ \bibinfo {pages}
  {074506} (\bibinfo {year} {2017})}\BibitemShut {NoStop}%
\bibitem [{\citenamefont {Sbalzarini}\ and\ \citenamefont
  {Koumoutsakos}(2005)}]{Sbalzarini2005}%
  \BibitemOpen
  \bibfield  {author} {\bibinfo {author} {\bibfnamefont {I.~F.}\ \bibnamefont
  {Sbalzarini}}\ and\ \bibinfo {author} {\bibfnamefont {P.}~\bibnamefont
  {Koumoutsakos}},\ }\bibfield  {title} {\enquote {\bibinfo {title} {Feature
  point tracking and trajectory analysis for video imaging in cell biology},}\
  }\href {\doibase https://doi.org/10.1016/j.jsb.2005.06.002} {\bibfield
  {journal} {\bibinfo  {journal} {J. Struct. Biol.}\ }\textbf {\bibinfo
  {volume} {151}},\ \bibinfo {pages} {182} (\bibinfo {year}
  {2005})}\BibitemShut {NoStop}%
\bibitem [{\citenamefont {{La Mantia}}\ \emph {et~al.}(2012)\citenamefont {{La
  Mantia}}, \citenamefont {Chagovets}, \citenamefont {Rotter},\ and\
  \citenamefont {Skrbek}}]{LaMantia2012}%
  \BibitemOpen
  \bibfield  {author} {\bibinfo {author} {\bibfnamefont {M.}~\bibnamefont {{La
  Mantia}}}, \bibinfo {author} {\bibfnamefont {T.~V.}\ \bibnamefont
  {Chagovets}}, \bibinfo {author} {\bibfnamefont {M.}~\bibnamefont {Rotter}}, \
  and\ \bibinfo {author} {\bibfnamefont {L.}~\bibnamefont {Skrbek}},\
  }\bibfield  {title} {\enquote {\bibinfo {title} {Testing the performance of a
  cryogenic visualization system on thermal counterflow by using hydrogen and
  deuterium solid tracers},}\ }\href {\doibase 10.1063/1.4719917} {\bibfield
  {journal} {\bibinfo  {journal} {Rev. Sci. Instrum.}\ }\textbf {\bibinfo
  {volume} {83}},\ \bibinfo {pages} {055109} (\bibinfo {year}
  {2012})}\BibitemShut {NoStop}%
\bibitem [{\citenamefont {Wang}\ \emph {et~al.}(1987)\citenamefont {Wang},
  \citenamefont {Swanson},\ and\ \citenamefont {Donnelly}}]{Wang1987}%
  \BibitemOpen
  \bibfield  {author} {\bibinfo {author} {\bibfnamefont {R.~T.}\ \bibnamefont
  {Wang}}, \bibinfo {author} {\bibfnamefont {C.~E.}\ \bibnamefont {Swanson}}, \
  and\ \bibinfo {author} {\bibfnamefont {R.~J.}\ \bibnamefont {Donnelly}},\
  }\bibfield  {title} {\enquote {\bibinfo {title} {Anisotropy and drift of a
  vortex tangle in helium {II}},}\ }\href {\doibase 10.1103/PhysRevB.36.5240}
  {\bibfield  {journal} {\bibinfo  {journal} {Phys. Rev. B}\ }\textbf {\bibinfo
  {volume} {36}},\ \bibinfo {pages} {5240} (\bibinfo {year}
  {1987})}\BibitemShut {NoStop}%
\bibitem [{\citenamefont {Donnelly}(1991)}]{Donnelly1991}%
  \BibitemOpen
  \bibfield  {author} {\bibinfo {author} {\bibfnamefont {R.~J.}\ \bibnamefont
  {Donnelly}},\ }\href@noop {} {\emph {\bibinfo {title} {Quantized vortices in
  helium {II}}}}\ (\bibinfo  {publisher} {Cambridge University Press},\
  \bibinfo {year} {1991})\BibitemShut {NoStop}%
\bibitem [{\citenamefont {Donnelly}\ and\ \citenamefont
  {Barenghi}(1998)}]{Donnelly1998}%
  \BibitemOpen
  \bibfield  {author} {\bibinfo {author} {\bibfnamefont {R.~J.}\ \bibnamefont
  {Donnelly}}\ and\ \bibinfo {author} {\bibfnamefont {C.~F.}\ \bibnamefont
  {Barenghi}},\ }\bibfield  {title} {\enquote {\bibinfo {title} {The observed
  properties of liquid helium at the saturated vapor pressure},}\ }\href
  {\doibase 10.1063/1.556028} {\bibfield  {journal} {\bibinfo  {journal} {J.
  Phys. Chem. Ref. Data}\ }\textbf {\bibinfo {volume} {27}},\ \bibinfo {pages}
  {1217} (\bibinfo {year} {1998})}\BibitemShut {NoStop}%
\bibitem [{\citenamefont {La~Mantia}\ \emph {et~al.}(2013)\citenamefont
  {La~Mantia}, \citenamefont {Duda}, \citenamefont {Rotter},\ and\
  \citenamefont {Skrbek}}]{LaMantia2013}%
  \BibitemOpen
  \bibfield  {author} {\bibinfo {author} {\bibfnamefont {M.}~\bibnamefont
  {La~Mantia}}, \bibinfo {author} {\bibfnamefont {D.}~\bibnamefont {Duda}},
  \bibinfo {author} {\bibfnamefont {M.}~\bibnamefont {Rotter}}, \ and\ \bibinfo
  {author} {\bibfnamefont {L.}~\bibnamefont {Skrbek}},\ }\bibfield  {title}
  {\enquote {\bibinfo {title} {Lagrangian accelerations of particles in
  superfluid turbulence},}\ }\href {\doibase 10.1017/jfm.2013.31} {\bibfield
  {journal} {\bibinfo  {journal} {J. Fluid Mech.}\ }\textbf {\bibinfo {volume}
  {717}},\ \bibinfo {pages} {R9} (\bibinfo {year} {2013})}\BibitemShut
  {NoStop}%
\bibitem [{\citenamefont {La~Mantia}\ \emph {et~al.}(2016)\citenamefont
  {La~Mantia}, \citenamefont {\ifmmode \check{S}\else
  \v{S}\fi{}van\ifmmode~\check{c}\else \v{c}\fi{}ara}, \citenamefont {Duda},\
  and\ \citenamefont {Skrbek}}]{LaMantia2016b}%
  \BibitemOpen
  \bibfield  {author} {\bibinfo {author} {\bibfnamefont {M.}~\bibnamefont
  {La~Mantia}}, \bibinfo {author} {\bibfnamefont {P.}~\bibnamefont {\ifmmode
  \check{S}\else \v{S}\fi{}van\ifmmode~\check{c}\else \v{c}\fi{}ara}}, \bibinfo
  {author} {\bibfnamefont {D.}~\bibnamefont {Duda}}, \ and\ \bibinfo {author}
  {\bibfnamefont {L.}~\bibnamefont {Skrbek}},\ }\bibfield  {title} {\enquote
  {\bibinfo {title} {Small-scale universality of particle dynamics in quantum
  turbulence},}\ }\href {\doibase 10.1103/PhysRevB.94.184512} {\bibfield
  {journal} {\bibinfo  {journal} {Phys. Rev. B}\ }\textbf {\bibinfo {volume}
  {94}},\ \bibinfo {pages} {184512} (\bibinfo {year} {2016})}\BibitemShut
  {NoStop}%
\bibitem [{\citenamefont {Gao}\ \emph {et~al.}(2016{\natexlab{b}})\citenamefont
  {Gao}, \citenamefont {Guo}, \citenamefont {L'vov}, \citenamefont {Pomyalov},
  \citenamefont {Skrbek}, \citenamefont {Varga},\ and\ \citenamefont
  {Vinen}}]{Gao2015JETP}%
  \BibitemOpen
  \bibfield  {author} {\bibinfo {author} {\bibfnamefont {J.}~\bibnamefont
  {Gao}}, \bibinfo {author} {\bibfnamefont {W.}~\bibnamefont {Guo}}, \bibinfo
  {author} {\bibfnamefont {V.S.}\ \bibnamefont {L'vov}}, \bibinfo {author}
  {\bibfnamefont {A.}~\bibnamefont {Pomyalov}}, \bibinfo {author}
  {\bibfnamefont {L.}~\bibnamefont {Skrbek}}, \bibinfo {author} {\bibfnamefont
  {E.}~\bibnamefont {Varga}}, \ and\ \bibinfo {author} {\bibfnamefont {W.F.}\
  \bibnamefont {Vinen}},\ }\bibfield  {title} {\enquote {\bibinfo {title} {The
  decay of counterflow turbulence in superfluid \textsuperscript{4}{He}},}\
  }\href {\doibase 10.7868/S0370274X16100088} {\bibfield  {journal} {\bibinfo
  {journal} {JETP Lett.}\ }\textbf {\bibinfo {volume} {103}},\ \bibinfo {pages}
  {732} (\bibinfo {year} {2016}{\natexlab{b}})}\BibitemShut {NoStop}%
\bibitem [{\citenamefont {Kivotides}\ \emph {et~al.}(2007)\citenamefont
  {Kivotides}, \citenamefont {Barenghi},\ and\ \citenamefont
  {Sergeev}}]{Kivotides2007}%
  \BibitemOpen
  \bibfield  {author} {\bibinfo {author} {\bibfnamefont {D.}~\bibnamefont
  {Kivotides}}, \bibinfo {author} {\bibfnamefont {C.~F.}\ \bibnamefont
  {Barenghi}}, \ and\ \bibinfo {author} {\bibfnamefont {Y.~A.}\ \bibnamefont
  {Sergeev}},\ }\bibfield  {title} {\enquote {\bibinfo {title} {Collision of a
  tracer particle and a quantized vortex in superfluid helium: Self-consistent
  calculations},}\ }\href {\doibase 10.1103/PhysRevB.75.212502} {\bibfield
  {journal} {\bibinfo  {journal} {Phys. Rev. B}\ }\textbf {\bibinfo {volume}
  {75}},\ \bibinfo {pages} {212502} (\bibinfo {year} {2007})}\BibitemShut
  {NoStop}%
\bibitem [{\citenamefont {Vinen}()}]{VinenPrivate}%
  \BibitemOpen
  \bibfield  {author} {\bibinfo {author} {\bibfnamefont {W.~F.}\ \bibnamefont
  {Vinen}},\ }\href@noop {} {}\bibinfo {howpublished} {Private
  communication}\BibitemShut {NoStop}%
\bibitem [{\citenamefont {Vinen}\ and\ \citenamefont
  {Niemela}(2002)}]{Vinen2002}%
  \BibitemOpen
  \bibfield  {author} {\bibinfo {author} {\bibfnamefont {W.~F.}\ \bibnamefont
  {Vinen}}\ and\ \bibinfo {author} {\bibfnamefont {J.~J.}\ \bibnamefont
  {Niemela}},\ }\bibfield  {title} {\enquote {\bibinfo {title} {Quantum
  turbulence},}\ }\href {\doibase 10.1023/A:1019695418590} {\bibfield
  {journal} {\bibinfo  {journal} {J. Low Temp. Phys.}\ }\textbf {\bibinfo
  {volume} {128}},\ \bibinfo {pages} {167} (\bibinfo {year}
  {2002})}\BibitemShut {NoStop}%
\bibitem [{\citenamefont {Gao}\ \emph {et~al.}()\citenamefont {Gao},
  \citenamefont {Guo}, \citenamefont {Yui}, \citenamefont {Tsubota},\ and\
  \citenamefont {Vinen}}]{GaoArxiv}%
  \BibitemOpen
  \bibfield  {author} {\bibinfo {author} {\bibfnamefont {J.}~\bibnamefont
  {Gao}}, \bibinfo {author} {\bibfnamefont {W.}~\bibnamefont {Guo}}, \bibinfo
  {author} {\bibfnamefont {S.}~\bibnamefont {Yui}}, \bibinfo {author}
  {\bibfnamefont {M.}~\bibnamefont {Tsubota}}, \ and\ \bibinfo {author}
  {\bibfnamefont {W.~F.}\ \bibnamefont {Vinen}},\ }\href@noop {} {\enquote
  {\bibinfo {title} {Dissipation in quantum turbulence in superfluid
  \textsuperscript{4}{He} above {1K}},}\ }\bibinfo {note} {ArXiv:1804.01655
  [cond-mat.other]}\BibitemShut {NoStop}%
\bibitem [{\citenamefont {Schwarz}(1988)}]{Schwarz1988}%
  \BibitemOpen
  \bibfield  {author} {\bibinfo {author} {\bibfnamefont {K.~W.}\ \bibnamefont
  {Schwarz}},\ }\bibfield  {title} {\enquote {\bibinfo {title}
  {Three-dimensional vortex dynamics in superfluid $^{4}\mathrm{He}$:
  Homogeneous superfluid turbulence},}\ }\href {\doibase
  10.1103/PhysRevB.38.2398} {\bibfield  {journal} {\bibinfo  {journal} {Phys.
  Rev. B}\ }\textbf {\bibinfo {volume} {38}},\ \bibinfo {pages} {2398}
  (\bibinfo {year} {1988})}\BibitemShut {NoStop}%
\bibitem [{\citenamefont {Xu}\ \emph {et~al.}(2007)\citenamefont {Xu},
  \citenamefont {Ouellette},\ and\ \citenamefont {Bodenschatz}}]{HXu2007}%
  \BibitemOpen
  \bibfield  {author} {\bibinfo {author} {\bibfnamefont {H.}~\bibnamefont
  {Xu}}, \bibinfo {author} {\bibfnamefont {N.~T.}\ \bibnamefont {Ouellette}}, \
  and\ \bibinfo {author} {\bibfnamefont {E.}~\bibnamefont {Bodenschatz}},\
  }\bibfield  {title} {\enquote {\bibinfo {title} {Curvature of {Lagrangian}
  trajectories in turbulence},}\ }\href {\doibase
  10.1103/PhysRevLett.98.050201} {\bibfield  {journal} {\bibinfo  {journal}
  {Phys. Rev. Lett.}\ }\textbf {\bibinfo {volume} {98}},\ \bibinfo {pages}
  {050201} (\bibinfo {year} {2007})}\BibitemShut {NoStop}%
\bibitem [{\citenamefont {Burov}\ \emph {et~al.}(2013)\citenamefont {Burov},
  \citenamefont {Tabei}, \citenamefont {Huynh}, \citenamefont {Murrell},
  \citenamefont {Philipson}, \citenamefont {Rice}, \citenamefont {Gardel},
  \citenamefont {Scherer},\ and\ \citenamefont {Dinner}}]{Burov2013}%
  \BibitemOpen
  \bibfield  {author} {\bibinfo {author} {\bibfnamefont {S.}~\bibnamefont
  {Burov}}, \bibinfo {author} {\bibfnamefont {S.~M.~A.}\ \bibnamefont {Tabei}},
  \bibinfo {author} {\bibfnamefont {T.}~\bibnamefont {Huynh}}, \bibinfo
  {author} {\bibfnamefont {M.~P.}\ \bibnamefont {Murrell}}, \bibinfo {author}
  {\bibfnamefont {L.~H.}\ \bibnamefont {Philipson}}, \bibinfo {author}
  {\bibfnamefont {S.~A.}\ \bibnamefont {Rice}}, \bibinfo {author}
  {\bibfnamefont {M.~L.}\ \bibnamefont {Gardel}}, \bibinfo {author}
  {\bibfnamefont {N.~F.}\ \bibnamefont {Scherer}}, \ and\ \bibinfo {author}
  {\bibfnamefont {A.~R.}\ \bibnamefont {Dinner}},\ }\bibfield  {title}
  {\enquote {\bibinfo {title} {Distribution of directional change as a
  signature of complex dynamics},}\ }\href {\doibase 10.1073/pnas.1319473110}
  {\bibfield  {journal} {\bibinfo  {journal} {Proc. Natl. Acad. Sci. USA}\
  }\textbf {\bibinfo {volume} {110}},\ \bibinfo {pages} {19689} (\bibinfo
  {year} {2013})}\BibitemShut {NoStop}%
\bibitem [{\citenamefont {Bos}\ \emph {et~al.}(2015)\citenamefont {Bos},
  \citenamefont {Kadoch},\ and\ \citenamefont {Schneider}}]{Bos2015}%
  \BibitemOpen
  \bibfield  {author} {\bibinfo {author} {\bibfnamefont {W.~J.~T.}\
  \bibnamefont {Bos}}, \bibinfo {author} {\bibfnamefont {B.}~\bibnamefont
  {Kadoch}}, \ and\ \bibinfo {author} {\bibfnamefont {K.}~\bibnamefont
  {Schneider}},\ }\bibfield  {title} {\enquote {\bibinfo {title} {Angular
  statistics of {Lagrangian} trajectories in turbulence},}\ }\href {\doibase
  10.1103/PhysRevLett.114.214502} {\bibfield  {journal} {\bibinfo  {journal}
  {Phys. Rev. Lett.}\ }\textbf {\bibinfo {volume} {114}},\ \bibinfo {pages}
  {214502} (\bibinfo {year} {2015})}\BibitemShut {NoStop}%
\bibitem [{\citenamefont {Bhatnagar}\ \emph {et~al.}(2016)\citenamefont
  {Bhatnagar}, \citenamefont {Gupta}, \citenamefont {Mitra}, \citenamefont
  {Perlekar}, \citenamefont {Wilkinson},\ and\ \citenamefont
  {Pandit}}]{Bhatnagar2016}%
  \BibitemOpen
  \bibfield  {author} {\bibinfo {author} {\bibfnamefont {A.}~\bibnamefont
  {Bhatnagar}}, \bibinfo {author} {\bibfnamefont {A.}~\bibnamefont {Gupta}},
  \bibinfo {author} {\bibfnamefont {D.}~\bibnamefont {Mitra}}, \bibinfo
  {author} {\bibfnamefont {P.}~\bibnamefont {Perlekar}}, \bibinfo {author}
  {\bibfnamefont {M.}~\bibnamefont {Wilkinson}}, \ and\ \bibinfo {author}
  {\bibfnamefont {R.}~\bibnamefont {Pandit}},\ }\bibfield  {title} {\enquote
  {\bibinfo {title} {Deviation-angle and trajectory statistics for inertial
  particles in turbulence},}\ }\href {\doibase 10.1103/PhysRevE.94.063112}
  {\bibfield  {journal} {\bibinfo  {journal} {Phys. Rev. E}\ }\textbf {\bibinfo
  {volume} {94}},\ \bibinfo {pages} {063112} (\bibinfo {year}
  {2016})}\BibitemShut {NoStop}%
\bibitem [{\citenamefont {Barenghi}()}]{BarenghiPrivate}%
  \BibitemOpen
  \bibfield  {author} {\bibinfo {author} {\bibfnamefont {C.~F.}\ \bibnamefont
  {Barenghi}},\ }\href@noop {} {}\bibinfo {howpublished} {Private
  communication}\BibitemShut {NoStop}%
\end{thebibliography}
%

\end{document}